\documentclass[useAMS]{mn2e}
\usepackage{amssymb,amsmath,times,graphicx,psfig}
\usepackage{mathtools}
\usepackage{color}
\usepackage{epstopdf}
\def\gsim{ \lower .75ex \hbox{$\sim$} \llap{\raise .27ex \hbox{$>$}} }
\def\lsim{ \lower .75ex\hbox{$\sim$} \llap{\raise .27ex \hbox{$<$}} }

\def\sc{Schwarzschild}
\def\beq{\begin{equation}}
\def\eeq{\end{equation}}

\def\sc{Schwarzschild}

\def\g2av{\langle\gamma^2\rangle}

\title[Radiative acceleration in spine--sheath blazar jets]{On radiative acceleration in spine--sheath structured blazar jets}

\author[A. Chhotray et al.]
{A. Chhotray$^1$\thanks{Email: chhotraa@oregonstate.edu}, 
F. Nappo$^{2,3}$, G. Ghisellini$^3$, O. S. Salafia$^{4,3}$, F. Tavecchio$^3$, D. Lazzati$^{^1}$ 
\\
$^1$ Department of Physics, Oregon State University, 97331 Corvallis, Oregon, USA \\
$^2$ Dipartimento di Scienza e Alta Tecnologia, Universit\`a degli Studi dell'Insubria, Via Valleggio 11, I--22100 Como, Italy\\
$^3$ INAF -- Osservatorio Astronomico di Brera, Via Bianchi 46, I--23807 Merate, Italy\\
$^4$ Dipartimento di Fisica ``G. Occhialini", Universit\`a degli Studi di Milano--Bicocca, P.za della Scienza 3, I--20126 Milano, Italy\\ 
}
\begin{document}  

\maketitle

\begin{abstract}
  It has been proposed that blazar jets are structured, with a
  fast spine surrounded by a slower sheath or layer. This structured jet model explains some properties of their emission and morphology. Because
  of their relative motion, the radiation produced by one component is
  seen amplified by the other, thus enhancing the inverse
  Compton emission of both. Radiation is emitted anisotropically in
  the comoving frames, and causes the emitting plasma to recoil.  As
  seen in the observer frame, this corresponds to a deceleration of
  the fastest component (the spine) and an acceleration of the slower
  one (the layer). 
  While the deceleration of the spine has already been 
  investigated, here we study for the first time the acceleration of the 
  sheath and find self--consistent velocity profile solutions for both the 
  spine and the sheath while accounting for radiative cooling. 
  We find that the sheath can be accelerated to the velocities required  
  by the observations if its leptons remain energetic in the acceleration region, assumed to be of the order of  $\sim$100 \sc\ radii, 
%   long enough time (how can we better quantify this?) \ remain hot?, 
  demanding continuous injection of energetic particles in that region.
% We explore a large region of the parameter space we identify/find cases 
% where  structured jets can achieve Lorentz factors consistent with observations
% via radiative acceleration. 
% the sheath 
% AC
%
\end{abstract}

\section{Introduction}
%\begin{itemize}
%\item  History
%
%\item enhancement of Compton: good for blazars  may be useful for GRB?
%
%\item deceleration of the spine in GG Tav
%
%\item here acceleration of the layer 
%
%\item here feedback 
%
%\item we need simplifications to solve the problem 
%
%\item Concepts of ``Compton drag" and ``Compton rocket"
%
%\end{itemize}
Relativistic jets in low-power radio--loud active galactic nuclei (AGN) are thought to be
%Relativistic jets in radio--loud quasars and radio--galaxies are thought to be 
structured, namely composed of a fast central
part, that we call the spine, and a sheath or a layer surrounding it,
moving at a slower speed.  There are several arguments that support
the structured jet hypothesis. It is very unlikely that the jet plasma
moves with a large bulk Lorentz factor $\Gamma$ ($\sim$ 10--15) inside
the jet and with $\Gamma=1$ just outside it.  The velocity of the
plasma should decrease gradually across the edge of the jet because of
shear viscosity and/or Kelvin--Helmoltz instabilities (e.g., Henri \&
Pelletier 1991; for a review see Ferrari 1998; see also Bodo et al. 2003). 
Structured jets could also result from the acceleration mechanism itself (e.g., McKinney 2006). 
% Komissarov (1990) and Laing (1993) pointed out some
% observational consequence of the structured jet scenario {\bf..Do we mention these consequences anywhere?}

Observationally, the emission of high energy $\gamma$--ray radiation
requires a large bulk Lorentz factor in blazars (i.e., sources whose
jets are pointing at us), to avoid suppression by the
$\gamma\gamma \to$ e$^\pm$ process.  Low-power, TeV emitting BL Lacs require the
largest values of $\Gamma$ among all blazars (Tavecchio et al. 2001,
Kino, Takahara \& Kusunose 2002; Katarzynski, Sol \& Kus 2003;
Krawczynski, Coppi \& Aharonian 2002; Konopelko et al. 2003; Tavecchio
et al., 2010).  However, the Large Area Telescope (LAT) onboard the
{\it Fermi} satellite has detected (low-power) radio galaxies at $\sim$ GeV
energies (Abdo et al. 2010; Grandi 2012).  Their radiation cannot be
the de--beamed emission coming from plasma moving with
$\Gamma\sim 15$, since the de--beaming would be too strong, making the
flux undetectable.  The GeV radiation of radio galaxies must be
produced by material moving with $\Gamma \sim 3$ (Ghisellini,
Tavecchio \& Chiaberge 2005) which is high enough to avoid the $\gamma\gamma$
absorption process but sufficiently small to avoid strong de--beaming
of the flux.

Detailed VLBI radio maps of Mkn 501 revealed a {\it limb brightening}
morphology, interpreted as evidence of a slower external flow
surrounding a faster spine (Giroletti et al. 2004).  Similar results
have been obtained for Mkn 421 (Giroletti et al. 2006), 0502+675 and
1722+119 (Piner \& Edwards 2014).

In addition to the above evidences for structured AGN jets,
there is also mounting evidence for a decelerating spine in TeV BL
Lacs, and therefore radial structure.  
Many TeV BL Lacs are not
superluminal sources at the $\sim$pc scale (Edwards \& Piner 2002;
Piner \& Edwards 2004, 2014; Piner, Pant \& Edwards 2010, 2008) even
though they require the highest bulk Lorentz factors in the TeV
emitting region (that is in most, but not all, cases located at
sub--pc distances from the black hole).

Georganopoulos \& Kazanas (2003) proposed a model in which the jet 
% is structured along the radial direction{\bf(?)}, with  
has a fast inner part and a slower part further out.
%
%rapidly decelerating in the $\gamma$--ray zone {\bf (@ GG - after reading their paper 
%	I don't think they discuss any radial structure in their paper - they are 
%	talking about a decelerating jet along the z direction with two zones - the fast zone close to the base and the slower zone further out - AC)}.  
In their model, the fast base of the jet sees the radiation produced by the slower  
zone relativistically boosted.  
Analogously, the slow part of the jet
sees the radiation coming from the fast base of the jet
relativistically boosted.  
The radiation energy density seen by both
components is amplified with respect to the pure one--zone model.

Ghisellini, Tavecchio \& Chiaberge (2005) proposed a
``spine--layer" (or spine--sheath) jet structure with the two
components having different velocities (the spine is faster). 
% {\bf the jet structured as a function of the angular distance from the jet axis.  }
As before, each component receives increased amounts of seed photons. 
In this configuration the fast spine could be decelerated by the Compton rocket 
effect (O'Dell 1981), justifying the decelerating jet model of Georganopoulos \& Kazanas (2003).  
The spine--layer model has been successfully applied to explain the high 
energy emission of
radiogalaxies (M87: Tavecchio \& Ghisellini 2008; NGC 1275: Tavecchio
\& Ghisellini, 2014; 3C 66B: Tavecchio \& Ghisellini 2009) and
slightly misaligned blazars (PKS 0521--36: D'Ammando et al. 2015). It
has also been shown to help the production of high energy neutrinos in
the relativistic jet of radio--sources (Tavecchio, Ghisellini \&
Guetta 2014).

In the original model and in the later application to specific sources, the
velocity of the layer was a free parameter, and was assumed to be constant.
On the other hand, the emitting plasma of the layer, being illuminated by the photons
of the spine, emits anisotropically in its comoving frame and thus must recoil.
The relative bulk Lorentz factor between the two structures is therefore bound
to decrease, limiting the seed amplification effect leading to the extra inverse
Compton emission.

%The aim of this paper is to study self-consistently the
%photon-mediated interaction between the two jet components that move
%with high relative velocity. 
%In particular, we aim to explore and describe 
%the dynamic coupling of the two radial components and to better understand this 
%physical feedback process, as it could be important for relativistic jets in
%general, including Gamma Ray Bursts (e.g., Rossi et al. 2002; Lazzati
%\& Begelman 2005).  

In the initial jet zone (where there is no radiative interplay between the spine and the layer), the jet launching mechanism could itself be responsible for  accelerating both the spine and the layer (e.g. McKinney 2006). An alternative option is that this launching mechanism is responsible only 
for the acceleration of the spine, while the layer gets accelerated radiatively. The aim of this paper is to study self-consistently the
photon-mediated interaction between the two jet components that move
with high relative velocity and thus find out which of the two options is preferred. In particular, we aim to explore and describe 
the dynamic coupling of the two radial components and to better understand this physical feedback process, as it could be important for relativistic jets in general, including Gamma Ray Bursts (e.g., Rossi et al. 2002; Lazzati
\& Begelman 2005).
%{\bf
%The same (? - same as the jet launching process?) acceleration process could be responsible for the 
%bulk Lorentz factor of both the spine and the sheath (e.g. McKinney 2006), in the
%initial jet zone where there is no radiative interplay between  them.
%Alternatively, the acceleration process could be responsible only 
%for the acceleration of the spine, while the sheath is accelerated radiatively.
%Our study aims to find out which of the two options is preferred.
%}

This paper is organized as follows: in \S~\ref{sec:setup} we discuss
the setup of the model and the assumption made to make it
mathematically tractable.  In \S~\ref{sec:res} we present and discuss our
results, and this is followed by our conclusions in \S~\ref{sec:discnconclu}. 
We divide the study in five parts: 
i) we find the velocity profile 
of the layer assuming a constant spine velocity, and assuming a fixed 
energy distribution of the emitting electrons in the spine--layer system; 
ii) we study the layer motion assuming that the emitting electrons are 
injected at the start, and then radiatively cool; 
iii) we explore the layer's motion for a case where the injected electrons maintain a fixed 
energy distribution inside a fixed volume (this fixed volume is discussed in greater detail 
in \S~\ref{sec:assump}) and cool radiatively once the plasma exits that volume; 
iv) we self--consistently calculate the motion of the spine and the layer,
under their reciprocal radiative influence; and 
v) we study how the self--consistent motion of the spine--layer system is
influenced by electron--positron pair loading.

\section{Set up of the model}\label{sec:setup}

\begin{figure} 
\vskip -0.6 cm
%\hskip -1.3 cm
%\psfig{file=layer.ps,width=13cm,height=12cm } 
\includegraphics[width=80mm]{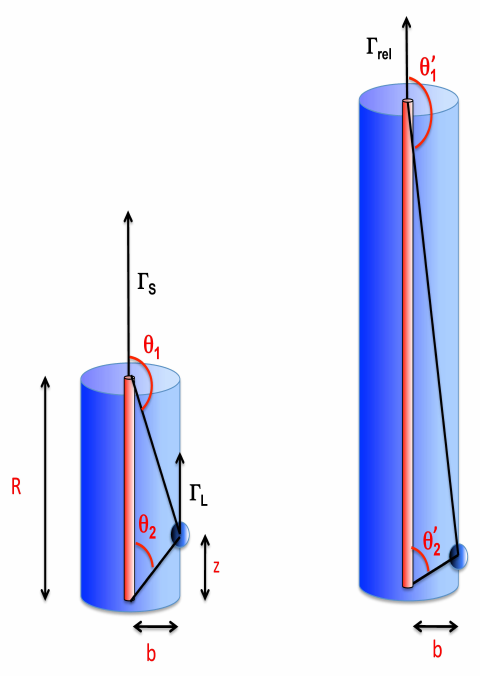}
\vskip -0.5 cm
\caption{
A cartoon depiction of the cylindrical spine-layer structure with its dimensions. 
On the left is the spine--layer jet as observed by a distant observer (such as on Earth) 
with both the layer and the spine in motion with Lorentz factors $\Gamma_{\rm L}$ 
and $\Gamma_{\rm S}$ respectively. 
The vertical height or size of the active region is $R \sim 10^{16}$ cm and the radial 
extent of the layer is $b \sim 10^{15}$ cm. 
We also show a layer particle at a height $z$ and the corresponding angles 
subtended at that height by the extremities of the active region. 
The figure on the right depicts the same structure as observed in the layer frame. 
For the layer the spine appears to move at a Lorentz factor $\Gamma'_{\rm rel}$ and 
relativistic effects elongate the vertical size of the active region. 
The radial width of the system however, is unaffected as it is orthogonal to the direction of motion.
}
\label{fig:cartoon}
\end{figure}

Our model consists of a cylindrical spine--layer 
structure as shown in Fig. \ref{fig:cartoon}.
%
% --------------------------------------
% -------------------------------------- 
This structure is a system of concentric cylinders with the spine being the inner cylindrical 
structure and the layer surrounds it. 
We use this model to describe the dynamical evolution of the spine and the 
layer due to Compton scattering by photons produced by the layer and the spine respectively.
%
%The discussions in this paper will focus on the physical quantities as measured in three different frames of reference, which are - 
%\begin{itemize}
%	\item Observer Frame - this is the frame of the distant observer and all quantities measured in this frame are unprimed
%	\item Spine frame - It is the frame of the moving spine and quantities measured in the frame of the spine are primed e.g. $L'_{\rm S}$
%	\item Sheath or Layer frame - This is the frame of the moving layer/sheath and the various quantities measured in this frame are also primed e.g. $L'_{\rm L}$
%\end{itemize}
The physical quantities of interest for investigating this problem are measured 
in three different reference frames. These three frames are:
% --------------------------------
\begin{enumerate}
\item The observer frame $K$: the quantities measured in this frame are identical to those measured by an 
observer on Earth, hence we will refer to this frame as the ``observer frame''. Any quantity measured in 
this frame will be marked as unprimed.

\item The layer frame $K'$: This is the frame instantaneously at rest with respect to the layer. 
It moves with respect to the observer frame at a variable Lorentz factor, denoted by $\Gamma_{\rm L}$. 
The quantities measured in this frame are marked with a single prime, e.g., $L'_{\rm S}$ is 
the spine luminosity as observed in the frame instantaneously co-moving with the layer.

\item The spine frame $K''$: This frame is co--moving with the spine plasma with a 
Lorentz factor $\Gamma_{\rm S}$.
The quantities measured in this frame are marked with a double prime, e.g., 
$L''_{\rm L}$ is the layer luminosity as measured in the spine frame.
\end{enumerate}
% --------------------------------

\subsection{Assumptions}\label{sec:assump}

We simplify the analysis of our spine--layer model by assuming that the spine 
is uni--dimensional and is in motion with an initial Lorentz factor 
$\Gamma_{S,0}$ (measured in the observer frame) along the jet--axis direction (referred to as $z$--axis)
as is depicted in the left panel of Fig. \ref{fig:cartoon} by the inner cylinder. 
The layer is the outer cylinder surrounding the spine, has a radius of $b \sim 10^{15}$ cm 
and, like the spine, travels along the jet axis with an initial Lorentz factor given by 
$\Gamma_{\rm L, 0}$ (subscript L denotes layer which we shall use synonymously with the sheath).

%The vertical dimension of the spine-sheath structure is given by 
%$R = 10^{16}$ cm as is depicted in the left panel of figure 1. 
%In our model, the spine is 'activated' by pinching of the jet.

In our model we assume that the both the spine and the layer are `active' 
only between two points which are fixed in the observer frame
and separated by a distance of $R=10^{16}$ cm,  
implying the emitting volume to be fixed in that frame.
Such an active region can be a result of a standing shock, 
where energy dissipation happens between fixed points.
% due to turbulence arising from jet pinching 
% \textbf{(\textit{INSERT REFERENCES on the spot I don't know -- GG})}.
Both the spine and the sheath emit radiation isotropically in their respective 
reference frames, however any other frame would observe the emissions to be beamed. We thus introduce the relativistic Doppler factor (hereafter {\it beaming} factor) 
$\delta$ as:
\begin{equation}
\delta_{\rm L}=\dfrac{1}{\Gamma_{\rm L}(1-\beta_{\rm L}\cos\theta)}
\label{eq:deltal}
\end{equation}
which is the beaming factor of the radiation produced in the layer frame 
as seen in the observer frame. 
$\theta$ is the angle between the jet axis and the line of sight as measured in the observer frame.
\begin{equation}
\delta_{\rm S}=\dfrac{1}{\Gamma_{\rm S}(1-\beta_{\rm S}\cos\theta)}
\label{eq:deltas}
\end{equation}
is the beaming factor of the radiation produced in the spine frame 
as seen in the observer frame, and 
\begin{equation}
\delta_{\rm S,L}=\dfrac{\delta_{\rm S}}{\delta_{\rm L}}
\label{eq:deltasl}
\end{equation}
is the relative beaming factor of the radiation produced in the spine 
frame and as seen in the layer frame (see also Georganopoulos \& Kazanas 2003; Ghisellini et al. 2005). 

The forces resulting from Compton scattering of the layer particles by the 
spine photons can drive/accelerate the sheath (in this work we consider scattering only 
within the Thomson regime). As the seed photons of Compton scattering are produced outside the layer, 
if the scattering particles are hot in $K'$, the scattered radiation is 
anisotropic also in the layer co--moving frame, making the layer recoil. 
For this reason, this interaction is often called {\it Compton Rocket} 
(Sikora et al., 1996; Ghisellini \& Tavecchio, 2010; Vuillaume, Henri \& Petrucci, 2015) and for hotter particles this driving force increases proportionally to their 
average internal energy $\langle\gamma^2\rangle$.
%{\it 
%We also ignore radiative losses due to synchrotron emissions.... 
%No, we assume a fixed particle distribution $N(\gamma)$. 
%This can be the result of injection and cooling. (GG). Also, we later  account  for cooling in the
%sense of a high energy cutoff, that resembles the fact that we stop to inject
%new particles, and the high energy ones die. I suggest to drop the sentence 
%"We also ignore radiative losses due to synchrotron emissions...." 

%We had originally invoked the presence of magnetic field along the jet axis to constrain 
%$b$, as otherwise the horizontal or radial component of the radiative force will result 
%in the lateral expansion of the jet. If we relax this assumption now how do we balance 
%the radial component of the force ?? - AC + @ GG - is this heavy enough?
%
%Well, the stability and general structure of the jet is outside the scope
%of the paper. Nobody knows for sure what collimates the normal jet, yet.
%One can say that magnetic field is responsible, but it is not a solved issue.
%--GG--
%}
% (as shown in the next section).

We assume that the layer particle is free to move in the direction parallel
to the jet axis. 
For simplicity, we assume that 
% A magnetic field along the jet axis constrains and fixes 
the distance between the layer and the jet axis, $b$, is fixed, despite
the presence of a radial radiative force, and thus also along the
normal to the jet axis. 
This can be achieved through the presence of a magnetic field.
% (- how do we justify this given that there is a radial component of the radiative force? - AC)}
%  \textbf{(\textit{NEED REFERENCE HERE!})}. 
To analyze the motion of the sheath we consider an infinitesimal part 
of the sheath at a position $z$ which we treat as an ``effective particle". 
The constituent particles inside the layer are representative 
of the sheath particles and the sheath can be thought of as composed 
of a collection of these effective particles (see also \S \ref{subsec:eqmotion}).

The right panel of Fig. \ref{fig:cartoon} depicts the structure 
as viewed in the frame of the sheath. 
The sheath finds the spine moving at a Lorentz factor 
$\Gamma_{\rm rel}=\Gamma_{\rm S}\Gamma_{\rm L}(1-\beta_{\rm S}\beta_{\rm L})$ and due to 
the aberration of light observes the vertical dimension of the active region to be larger than $R$. 
% Both the spine and the layer produce radiation 
%isotropically in their own respective reference frames, 
%however their emissions would be beamed in any other frame. 
We assume that the observer is located at a viewing angle of 
$\theta_{\rm view} = 5^{\circ}$.
%\begin{itemize}

%\item we need simplifications to solve the problem
%\item spine uni--dimensional
%\item layer as a ``bottle" at a distance $b$. 
%We assume that a toroidal magnetic field
%constrain  the motion in the direction parallel
%to the jet axis.
%\item we assume that the dissipation, and the production of the
%luminosity of both the spine and the layer, takes place between
%two points that are fixed in the frame of the black hole. 
%This implies that the emitting volume is fixed in that frame.
%The distance between these two points is $R$.
%\item The spine initially moves with $\Gamma_{\rm S}$
%and the layer initially moves with $\Gamma_{\rm L}$.
%\item luminosity of the spine and of the layer are .... in the frame....

%\item Doppler factors are ....
%\end{itemize}

\subsection{Particle distributions and cooling}
\label{sec:cool}

We assume that the particle distributions $N(\gamma)$ in the spine and the sheath 
to be a broken power law, with slope $p_1$ below and $p_2$ 
(where $p_2>p_1$) above the break $\gamma_{\rm b}$: 
% \begin{eqnarray}
% N(\gamma) \, &=& K \gamma^{-p_1} \qquad\qquad \,\,\,\,\,\, \gamma\le\gamma_{\rm b}
% \nonumber \\
% N(\gamma) \, &=& K \gamma_{\rm b}^{p_2-p_1} \gamma^{-p_2} \qquad\gamma>\gamma_{\rm b}
% \end{eqnarray}
\begin{equation}
N(\gamma) \, = \,
\begin{cases}
K \gamma^{-p_1} & \gamma_{\rm min}<\gamma\le\gamma_{\rm b} \\
K \gamma_{\rm b}^{p_2-p_1} \gamma^{-p_2} & \gamma_{\rm b}<\gamma<\gamma_{\rm max} \\
0 & {\rm otherwise}\\
\end{cases} \\
\label{eq:ngamma}
\end{equation}
where $\gamma_{\rm max}$ is the maximum Lorentz factor 
of the distribution that depends on the cooling rate (see \S \ref{sec:cool}).
For simplicity we omit hereafter the prime and the double prime for $N(\gamma)$ and $\gamma$.
%which 
We can use the distribution $N(\gamma)$ to calculate 
the averages $\langle \gamma\rangle$ and $\langle \gamma^2\rangle$ as: 
\begin{align}
 \langle \gamma\rangle &= \,\frac{\int N(\gamma)\gamma{\rm d}\gamma}{\int N(\gamma){\rm d}\gamma} \nonumber \\
 \langle \gamma^2\rangle &= \,\frac{\int N(\gamma)\gamma^2{\rm d}\gamma}{\int N(\gamma){\rm d}\gamma} \label{eq:gammav} 
 \end{align}
%
% \begin{eqnarray}
% \langle \gamma\rangle   \, &=& \frac{\frac{1}{p_1 - 2}\left(1 - \frac{1}{\gamma^{p_1 - 2}_b}\right) + \frac{\gamma^{p_2 - p_1}_b}{p_2 - 2}\left(\ \frac{1}{\gamma^{p_2-2}_b} - \frac{1}{\gamma^{p_2-2}_m} \right)}{\frac{1}{p_1 - 1}\left(1 - \frac{1}{\gamma^{p_1 - 1}_b}\right) + \frac{\gamma^{p_2 - p_1}_b}{p_2 - 1}\left(\ \frac{1}{\gamma^{p_2-1}_b} - \frac{1}{\gamma^{p_2-1}_m} \right)} \nonumber \\
% \langle \gamma^2\rangle \, &=& \frac{\frac{1}{p_1 - 3}\left(1 - \frac{1}{\gamma^{p_1 - 3}_b}\right) + \frac{\gamma^{p_2 - p_1}_b}{p_2 - 3}\left(\ \frac{1}{\gamma^{p_2-3}_b} - \frac{1}{\gamma^{p_2-3}_m} \right)}{\frac{1}{p_1 - 1}\left(1 - \frac{1}{\gamma^{p_1 - 1}_b}\right) + \frac{\gamma^{p_2 - p_1}_b}{p_2 - 1}\left(\ \frac{1}{\gamma^{p_2-1}_b} - \frac{1}{\gamma^{p_2-1}_m} \right)} 
% \end{eqnarray}
 %
 
%If $p_1 \le 2$ and $p_2 > 3$ the synchrotron $\nu F_{\nu}$ flux
%produced by the $N(\gamma)$ distribution is peaked.
%Thus the value of $\gamma_{\rm b}$ can be estimated from 
%the ratio of the peak frequencies of the ultraviolet spine synchrotron %emission, which also serves 
%as the seed photons for the GeV emission via inverse Compton from the layer.
%{\bf
%SENTENCE NOT CLEAR TO ME: from the ratio of ... what? -- GG -- Are we assuming %pure SSC?
%}
%
%\begin{equation}
%\gamma_{\rm b} = \left(\frac{3}{4}\frac{\nu_{\rm GeV}}{\nu_{\rm UV}}\right)^{1/2}
%\label{eq:gammabreak}
%\end{equation}
%
In our work we fix $\gamma_{\rm min}=1$, $p_1=2$ and $p_2=4$. A possible realisation of this case corresponds to continuous injection of electrons distributed as 
$Q(\gamma) \propto \gamma^{-s}$ above $\gamma_{\rm b}$,
and $Q(\gamma)=0$ below.
If radiative cooling is fast (i.e., even particles
with low Lorentz factors cool in a timescale shorter than the
light crossing time), the stationary $N(\gamma)$ distribution
will have a slope $p_2=s+1$ above $\gamma_{\rm b}$ and $p_1=2$ below.
%This case can be produced by a stationary system in which particles are continuously injected 
%for $\gamma > \gamma_{\rm b}$ with a power--law $Q(\gamma) \propto \gamma^{-s}$, where $s=3$. 
%For $\gamma < \gamma_{\rm b}$, the particles distribute according to the power--law index $p_1=2$, 
%because of radiative cooling due to synchrotron and Compton scattering. 
For $s=3$ (i.e., $p_2=4$) Eq. \ref{eq:gammav} gives:
\begin{align}\label{eq:aver2}
\langle \gamma\rangle   \, &=\, \frac{3}{2}\, \frac{2 \, 
{\rm ln}(\gamma_{\rm b})+1-(\gamma_{\rm b}/\gamma_{\rm max})^2}{3-2/
\gamma_{\rm b}-\gamma^2_{\rm b}/\gamma^3_{\rm max}} \nonumber \\
\langle \gamma^2\rangle \, &=\, 3\, \frac{2\, 
\gamma_{\rm b}-1-\gamma^2_{\rm b}/\gamma_{\rm max}}{3-2/\gamma_{\rm b}-\gamma^2_{\rm b}/
\gamma^3_{\rm max}} 
\end{align}
%

 %------------------------------------------------
\begin{figure} 
\includegraphics[width=80mm]{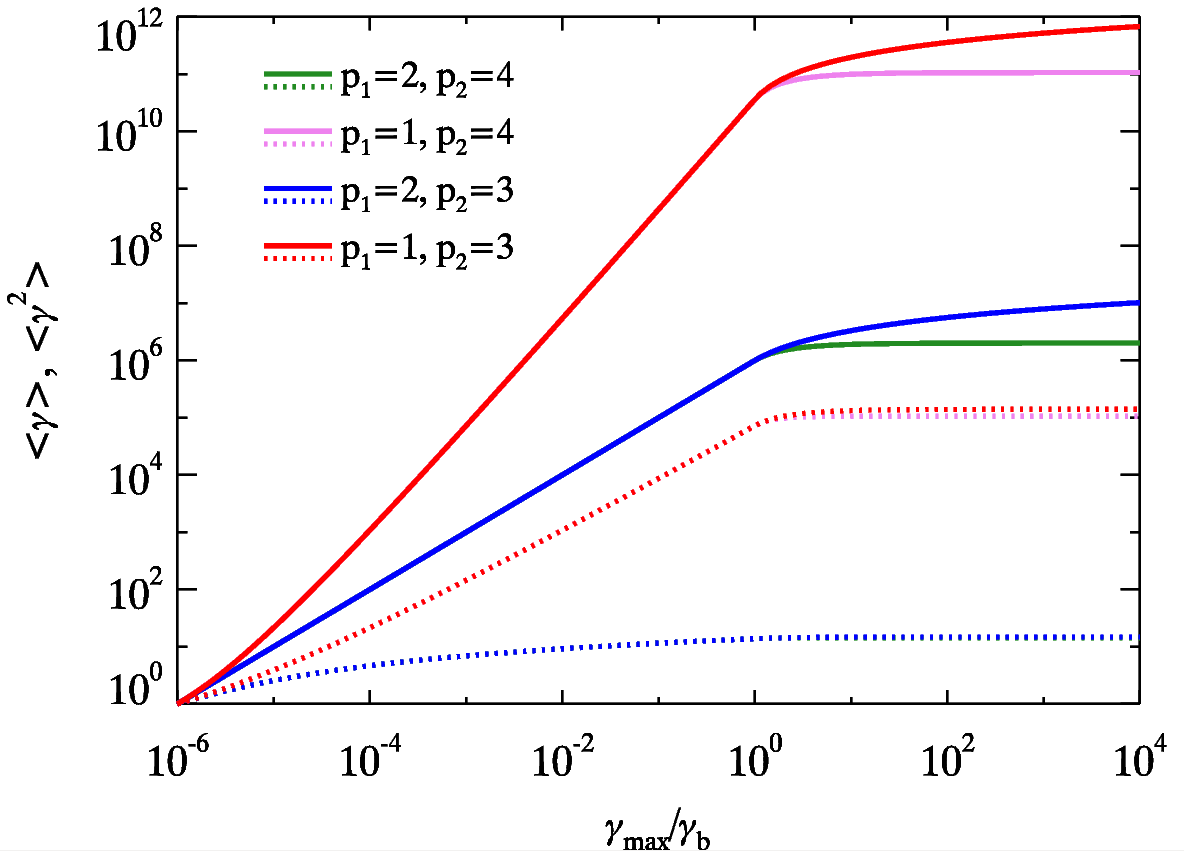}
\caption{
Evolution of $\langle\gamma\rangle$ (dotted lines)
and $\langle\gamma^2\rangle$ (solid lines) as a function 
of the ratio $\gamma_{\rm max}/\gamma_{\rm b}$ for different values of $p_1$ and $p_2$. 
As the hottest particles cool faster, the radiative cooling reduces the maximum Lorentz 
factor of particles: $\gamma_{\rm max}$. 
For our case of interest $p_1=2$ and $p_2=4$ (denoted by the green curve), 
when $\gamma_{\rm max}>\gamma_{\rm b}$ we observe that both $\langle\gamma\rangle$ and 
$\langle\gamma^2\rangle$ are constant. 
The averages start decreasing only when $\gamma_{\rm max}<\gamma_{\rm b}$. 
A very similar behavior is observed for other power law indices that 
are shown in the figure for comparison.}
\label{fig:average}
\end{figure}
%--------------------------------------- 
%%
%\begin{figure} 
%	\includegraphics[width=80mm]{Fig2.eps}
%	\caption{
%		EPS Evolution of $\langle\gamma\rangle$ (dotted lines)
%		and $\langle\gamma^2\rangle$ (solid lines) as a function 
%		of the ratio $\gamma_{\rm max}/\gamma_{\rm b}$ for different values of $p_1$ and $p_2$. 
%		As the hottest particles cool faster, the radiative cooling reduces the maximum Lorentz 
%		factor of particles: $\gamma_{\rm max}$. 
%		For our case of interest $p_1=2$ and $p_2=4$ (denoted by the green curve), 
%		when $\gamma_{\rm max}>\gamma_{\rm b}$ we observe that both $\langle\gamma\rangle$ and 
%		$\langle\gamma^2\rangle$ are constant. 
%		The averages start decreasing only when $\gamma_{\rm max}<\gamma_{\rm b}$. 
%		A very similar behavior is observed for other power law indices that 
%		are shown in the figure for comparison.}
%	\label{fig:average}
%\end{figure}
If the injection of particles is not continuous,  
the high energy particles are not replenished any longer, and
the $N(\gamma)$ distribution cuts--off at the cooling energy $\gamma_{\rm c}$,
which decreases with time.
The cooling of the plasma impacts the particle energy distribution 
which in turn affects the force that these particles experience (see Eq. \ref{eq:force}). 
Therefore we have to account for radiative cooling of the sheath/spine 
plasma due to irradiation by the spine/sheath photons. 
The cooling rate is (e.g. Rybicki \& Lightman 1979):
\beq
\dot{\gamma} = \frac{{\rm d} \gamma}{{\rm d}t'} = \frac{4}{3} \frac{\sigma_{\rm T} c U'%_{\rm L}
\gamma_{\rm max}^2 \beta_{\rm max}^2}{m_{\rm e} c^2} \label{eq:coolrate}
\eeq
where $ U'$ is the integrated radiation energy density 
in the layer frame,
$\gamma_{\rm max}$ and $\beta_{\rm max}$ 
are respectively the Lorentz factor and speed of the particle 
possessing the maximum internal energy 
(hence the subscript max) 
%{\bf (i.e. the cooling Lorentz factor)}
in a hot plasma.
At each timestep $\Delta t_i'$ we calculate the cooling Lorentz 
factor of the leptons using Eq. \ref{eq:coolrate}:
\begin{eqnarray}
\gamma_{\rm c, i} &=& \gamma_{\rm max, i-1} - \dot{\gamma}_{i-1} \Delta t'_i  \nonumber \\
&=& 
%\gamma_{\rm max, i-1} -{4 \sigma_{\rm T} c U'_{\rm i-1}
%\gamma^2_{\rm max, i-1} \beta^2_{\rm max,i-1} \over 3 m_{\rm e} c^2} \Delta t'_i
%\label{eq:gcool}
\gamma_{\rm max, i-1} - \frac{4}{3} \frac{\sigma_{\rm T} c U'_{\rm i-1} \gamma_{\rm max, i-1}^2 \beta_{\rm max, i-1}^2}{m_{\rm e} c^2} \Delta t'_i \label{eq:gcool}
\end{eqnarray} 
We assume that the particle distribution vanishes 
for $\gamma > \gamma_{\rm c}$: 
the cooling Lorentz factor $\gamma_{\rm c}$ becomes the new maximum Lorentz factor of the distribution, i.e.,    
$\gamma_{\rm max, i+1} = \gamma_{\rm c, i}$.
Since $\gamma_{\rm c}$ is time dependent, the averages 
$\langle \gamma \rangle$ and $\langle \gamma^2 \rangle$ (see eqs. \ref{eq:aver2}) 
also become time dependent. Graphically, the evolution of the averages as a function of the ratio $\gamma_{\rm max}/\gamma_{\rm b}$ (in other words, with time due to cooling) is depicted in the Fig.~\ref{fig:average} for several power-law indices.

\subsection{The equation of motion}
\label{subsec:eqmotion}

In order to study the trajectory of the sheath or the layer, we require the equation of motion given as
\begin{equation}
{{\rm d}p \over {\rm d}t} \, =\, F^\prime
\label{eq:force}
\end{equation}
where d$p$ and d$t$ are calculated in the same frame (any frame),
but $F^\prime$ is calculated in the frame {\it comoving with the particle}
(see e.g. Weinberg, 1972).
Since we assume that the layer and the spine are optically thin, we can calculate 
the motion of a single particle due to Compton scattering. 
We define $f$  as the ratio of number of leptons to protons and $f>1$ indicates the presence 
of pairs in the plasma. 
This enables us to study the motion of an ``effective particle'' of inertial mass 
$m_{\rm i}=m_{\rm p}/f+\langle \gamma \rangle m_{\rm e}$,
where the electron mass $m_e$ is multiplied by 
$\langle \gamma \rangle$ to account for the average 
internal energy of the leptons. Equation \ref{eq:force} can be written as:
\begin{equation}
m_{\rm i} c {{\rm d} (\Gamma\beta)  \over {\rm d}t} \, =\, F^\prime 
\label{eq:dgb}
\end{equation}
We begin by considering the motion of the layer due to the interaction with the 
radiation produced by the spine (moving with a constant bulk Lorentz factor $\Gamma_{\rm S}$).
In this case, the driving force $F^{\prime}_z (z)$ can be computed by considering the flux received by a particle in the layer located at a given $z$.
This flux will be produced at different heights of the spine, seen under
a different angle and with a different beaming.
Therefore we will have to integrate over the entire length of the spine while accounting for the different degrees of relativistic effects.\\
From the detailed calculations as shown in the appendix, we write the equation of motion 
(Eq. \ref{eq:dgb}) as:
\begin{equation}
{{\rm d} (\Gamma_{\rm L}\beta_{\rm L})  \over {\rm d}t} \, =  \, \frac{16}{9}{\sigma_{\rm T} \over m_{\rm i}bc^2} \langle \gamma^2 \rangle \eta \int_{\theta_1}^{\theta_2} \lambda^{\prime\prime}_{\rm S}  \dfrac{\delta^4_{\rm S}}{\delta^2_{\rm L}}\dfrac{\cos\theta-\beta_{\rm L}}{1-\beta_{\rm L}\cos\theta}{\rm d}\theta
\label{eq:motion_eq}
\end{equation}
where $\eta$ is a factor of the order of unity that depends on the geometry of 
the system (in this case we used $\eta=2/\pi$), 
$\lambda^{\prime\prime}_{\rm S}=\dfrac{{\rm d}L_{\rm S}^{\prime\prime}}{{\rm d}x^{\prime\prime}}$ 
is the spine comoving linear luminosity density which is connected to the spine 
isotropic luminosity $L_{\rm iso,S}$.
%\subsubsection{Computing the driving force}

\subsubsection{The drag Lorentz factor}\label{sec:drag}
This section introduces the physical meaning of drag Lorentz factor $\Gamma_{\rm L, drag}$ 
which we will frequently use to understand our results. 
$\Gamma_{\rm L, drag}$ is the value of $\Gamma_{\rm L}$ for which the $z$ component of 
the force as measured in the comoving layer frame vanishes. 
The net force acting on the effective layer particle at a certain position is computed by 
accounting for photons that hit the effective particle both from above and below its position. 
As seen in the observer frame $K$, photons that hit the sheath ``effective particle'' 
with an incident angle less than $1/\Gamma_{\rm L}$ with respect to the sheath's direction 
of motion appear to arrive from above the effective particle's position in the sheath comoving frame.
These photons decelerate the particle by imparting negative momentum (negative force). 
On the other hand, photons that are incident with an angle greater than $1/\Gamma_{\rm L}$ 
will accelerate the particle, generating a positive force.
The value of the layer Lorentz factor for which these positive and negative forces are 
equal is called the \textit{drag Lorentz factor}. 
Its value as a function of $z$ can be obtained by imposing the condition that 
$F^\prime_z=0$, and this simplifies Eq. \ref{eq:motion_eq} to: 
\begin{equation}
\int_{\theta_1}^{\theta_2} \delta^4_{\rm S}\Gamma^2_{\rm L,drag}
(1-\beta_{\rm L, drag}\cos\theta)(\cos\theta-\beta_{\rm L, drag}){\rm d}\theta = 0
\label{eq:gdrag}
\end{equation}
%
%{\bf $\lambda^{\prime\prime}_{\rm S}$ missing? - the drag value does not depend upon this quantity.}
Since the sign of the total force $F^\prime_z$ depends only on the sign 
of the integral in Eq. \ref{eq:motion_eq} or Eq. \ref{eq:gdrag}, we will have:
\begin{equation*}
\begin{cases}
F^\prime_z > 0 & {\rm if} \quad \Gamma_{\rm L}< \Gamma_{\rm L, drag}\\
F^\prime_z < 0 & {\rm if} \quad \Gamma_{\rm L}> \Gamma_{\rm L, drag}\\
F^\prime_z = 0 & {\rm if} \quad \Gamma_{\rm L}= \Gamma_{\rm L, drag}
\end{cases}
\end{equation*}

\subsection{Feedback}\label{subsec:feedb}

The previous subsections describe the dynamical evolution of the sheath plasma interacting 
via Compton scattering with the spine photons. 
However, we ignored the effect of the sheath photons on the spine to simplify the analysis. 
In this section, we relax that assumption by accounting for the 
interaction of the layer photons with the spine and how this interaction modifies 
the spine Lorentz factor $\Gamma_{\rm S}(z)$. 
As a result, we can explore how the feedback between the spine--layer structure self regulates its very own dynamical evolution. \\
To study the 
spine--layer feedback, we need to modify some equations of the previous section according to the following considerations:
% -------------------------------
\begin{itemize}

\item As the bulk Lorentz factors for both the spine and the layer can be modified, 
we must compute two profiles $\Gamma_{\rm S}(z)$ and $\Gamma_{\rm L}(z)$ 
for the spine and the layer respectively.

\item The linear luminosity density profiles of the spine and the layer  
($\lambda_{\rm S}$ and $\lambda_{\rm L}$) can vary with position, 
but we assume that the comoving luminosity density profiles are proportional to 
the average square of the particle energies constituting the emitting plasma, i.e., 
$\lambda^{\prime\prime}_{\rm S} \propto \langle \gamma^2_{\rm S} \rangle$ and 
$\lambda^{\prime}_{\rm L} \propto \langle \gamma^2_{\rm L} \rangle$. 
%It is $\langle\gamma^2\rangle$, so it is not the kinetic energy, but the average
%square of the particle energy...
Thus, if the internal energy content of the plasma changes by radiative cooling,
the comoving luminosity profile will no longer be independent of the position.

\item We take advantage of the symmetry of the problem and suppose that the effect 
of the spine on the layer and the inverse effect, i.e., of the layer 
on the spine can be expressed by the same relations by simply switching the subscripts.

\end{itemize} 
%-------------------------------

These considerations lead to the following system of 
differential equations describing the evolution of two 
``effective particles'' that represent the spine and
the layer (having masses 
$m_{\rm S}=m_{\rm p}/f_{\rm S}+\langle \gamma_{\rm S} \rangle m_{\rm e}$ 
and $m_{\rm L}=m_{\rm p}/f_{\rm L}+\langle \gamma_{\rm L} \rangle m_{\rm e}$ 
for the spine and the layer respectively):
\beq
\begin{dcases*}
 m_{\rm S} c \dfrac{{\rm d} (\Gamma_{\rm S}\beta_{\rm S})}{{\rm d}t}  
 \,= \, \dfrac{16}{9}\dfrac{\sigma_{\rm T} }{ bc} \langle \gamma^2_{\rm S} 
 \rangle \eta \int_{\theta_1}^{\theta_2} \lambda^{\prime}_{\rm L} 
 (\theta) \frac{\delta^4_{\rm L}}{\delta^2_{\rm S}}\dfrac{\cos\theta-\beta_{\rm S}}
 {1-\beta_{\rm S}\cos\theta}{\rm d}\theta  \\
 m_{\rm L} c \dfrac{{\rm d} (\Gamma_{\rm L}\beta_{\rm L})}{{\rm d}t}  
 \,= \, \dfrac{16}{9}\dfrac{\sigma_{\rm T} }{ bc} \langle \gamma^2_{\rm L} 
 \rangle \eta \int_{\theta_1}^{\theta_2} \lambda^{\prime\prime}_{\rm S} 
 (\theta) \dfrac{\delta^4_{\rm S}}{\delta^2_{\rm L}}\dfrac{\cos\theta-\beta_{\rm L}}
 {1-\beta_{\rm L}\cos\theta}{\rm d}\theta  
\end{dcases*} 
\label{eq:feedback}
\eeq
All the important steps leading to the above equations are fully described in the Appendix, 
along with the limits of integrations $\theta_1$, $\theta_2$ (which can be obtained using 
the Eqs.~\ref{eq:costh1} and \ref{eq:costh2} respectively). 
We can compute the Lorentz factor profiles $\Gamma_{\rm S}(z)$ and $\Gamma_{\rm L}(z)$ 
by numerically solving this system. The reader should note that the acceleration of the layer (spine) depends 
linearly on the product of the isotropic luminosity of the spine (layer) and the average 
of the square of the leptonic Lorentz factor 
of the layer (spine). We call this product $k$:
\begin{equation}
k_{\rm S}\, =\, L_{\rm iso, L}\cdot \langle \gamma^2_{\rm S} \rangle; \quad
k_{\rm L}\, =\, L_{\rm iso, S}\cdot \langle \gamma^2_{\rm L} \rangle
\end{equation}
Hence we treat this product as a single parameter.
\section{Results and Discussion}
\label{sec:res}

In this section we present and discuss the results of the numerical 
integration of the equations of motion considering different conditions:
\begin{itemize}
\item \S\ref{subsec:nocool_nofeed}:
Radiative acceleration of the layer: No cooling scenario %(\S \ref{subsec:nocool_nofeed});
\item \S \ref{cooling}: Radiative acceleration of the layer: Cooling scenario
\begin{itemize}
	\item \S\ref{subsec:cool_nofeed}:  Single injection %(\S \ref{subsec: coolsingleshock})
	\item \S\ref{subsec:continj}: Continuous injection
	%(\S \ref{subsec: cool_continuousshock})
\end{itemize}
\item \S \ref{subsec:feed}: The spine--layer feedback %(\S \ref{subsec:feed});
	\begin{itemize}
		\item \S\ref{subsec:feedback:continj}: Continuous injection with feedback
	\end{itemize}
\item \S \ref{subsec:pairsfeed}: The spine--layer feedback in e$^+$e$^-$ pair loaded plasmas
%\item (INTERMEDIATE CASE: Cooling out of $R$);
\end{itemize}

%\subsection{No cooling, no feedback}

\subsection{Radiative acceleration of the layer: No cooling scenario}
\label{subsec:nocool_nofeed}

In this section we start with the simplest and somewhat unrealistic scenario where: 
%there is no cooling of the sheath plasma and no feedback between the spine-sheath structure. 
%
\begin{itemize}

\item[i)] the spine moves with a constant bulk Lorentz factor $\Gamma_{\rm S}$ 
(not considering its deceleration due to layer photons);

\item[ii)] the sheath particle distribution does not change with time 
(i.e., the particles do not cool by radiative emission,
or the cooling is exactly compensated by injection of new particles).
%{\it This makes the case unphysical, as physically this situation implies the offset of 
%cooling by unrealistic conditions of continuous shocks/energetic particle 
%injection into the plasma. 
%Only then can the particles remain energized and maintain their distributions.}
\end{itemize}
By switching 'off' the cooling we have made this scenario 
somewhat unrealistic, but this simplifies the problem at hand and in turn allows us to gain greater insight and develop intuition about the radiative acceleration phenomena. 
% we argue that by not accounting for radiative cooling we have removed an 
% additional variable and thereby simplified the problem at hand. 
% We insist that by exploring this scenario we can gain greater insight and develop 
% intuition about the radiative acceleration phenomena. 
This will help to improve our understanding of more complex scenarios discussed later in the paper. We solve the equation of motion of the layer under different conditions 
(however we only vary a single parameter in each case to develop intuition) and we show 
the spatial profiles of $\left(\Gamma \beta\right)_{\rm L}$
and of the force perceived by the layer projected over the $z$--axis $F^\prime_z (z)$.

%------------------------------------------------
\begin{figure} 
\includegraphics[width=80mm]{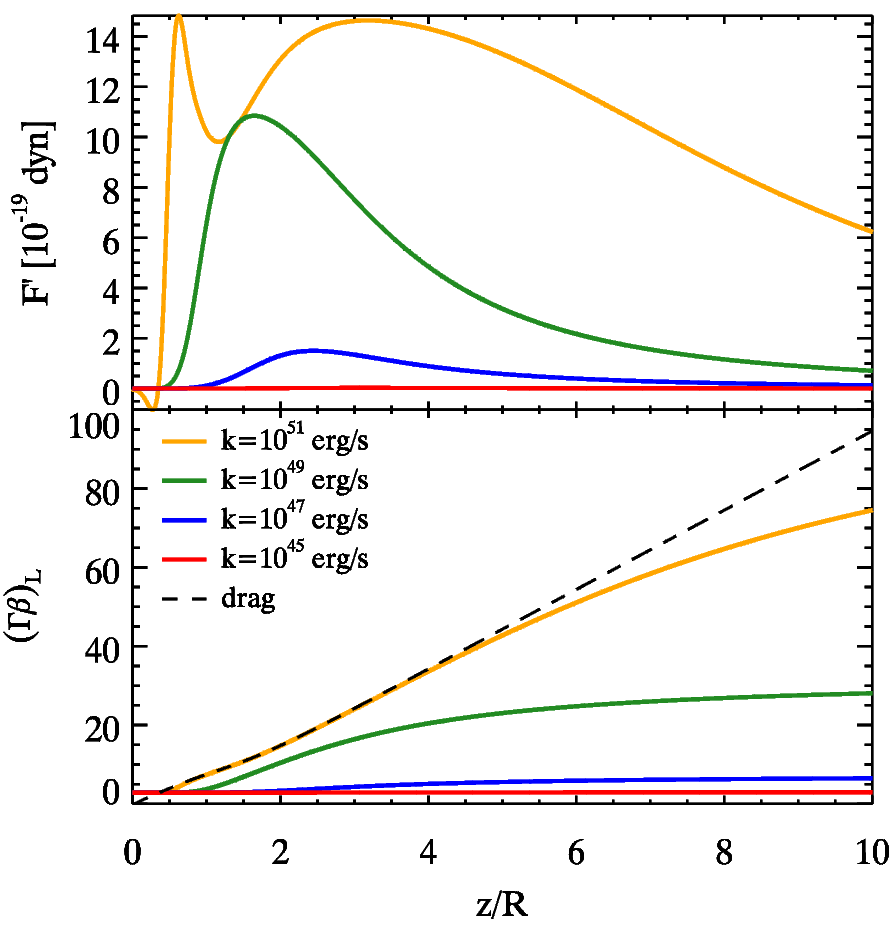}
\caption{
Radiative acceleration of the layer by varying  
$k_{\rm L}=L_{\rm iso,S}\cdot \langle \gamma^2 _{\rm L}\rangle$ 
% product of the spine isotropic luminosity and the average of the square of the leptonic Lorentz factor 
% of the layer $k=$ 
(values from $10^{45}$ to $10^{51}$ erg s$^{-1}$). 
The spine Lorentz factor is constant $\Gamma_{\rm S} = 15$. 
The initial Lorentz factor of the layer is $\Gamma_{\rm L,0}= 3$. 
The x--axis is common for the two panels and depicts the position $z$ normalized by 
the vertical structure dimension $R$. 
Top panel: radiative force $F^{\prime}$ as measured 
in the frame of the layer as a function of $z/R$.
Bottom panel: layer Lorentz factor $\left(\Gamma \beta\right)_{\rm L}$ (solid lines) 
and the drag Lorentz factor $\left(\Gamma \beta\right)_{\rm L, drag}$ 
of the layer (dashed line) as a function of $z/R$.
% {\bf I would show a inset with a zoom of the force from $z/R=0$ to 1... see below}
}
\label{fig:nocoolseqLg2}
\end{figure} 
%---------------------------------------
% \begin{figure}
% \includegraphics[width=80mm]{plot_zoom_fig3.eps}
% \caption{\textbf{Plot with zoomed inset - not very illuminating 
% according to both Francesco and me and so we would prefer to get rid of it altogether.} 
%   OK... --GG--
%
%
%	Radiative acceleration of the layer varying  
%	$k_{\rm L}=L_{\rm iso,S}\cdot \langle \gamma^2 \rangle$ 
%	% product of the spine isotropic luminosity and the average of the square of the leptonic Lorentz factor 
%	% of the layer $k=$ 
%	(values from $10^{45}$ to $10^{51}$ erg s$^{-1}$). 
%	The spine Lorentz factor is constant $\Gamma_{\rm S} = 15$. 
%	The initial Lorentz factor of the layer is $\Gamma_{\rm L,0}= 3$. 
%	Top panel: radiative force $F^{\prime}$ as measured 
%	in the frame of the layer as a function of $z$ measured in 
%	unit of the spine length $R$. 
%	Bottom panel: layer Lorentz factor $\left(\Gamma \beta\right)_{\rm L}$ (solid lines) 
%	and the drag Lorentz factor $\left(\Gamma \beta\right)_{\rm L, drag}$ 
%	of the layer (dashed line) as a function of $z$.}
%\label{fig:nocoolseqLg2zoom}
%\end{figure} 
%---------------------------------------
\begin{figure} 
\includegraphics[width=80mm]{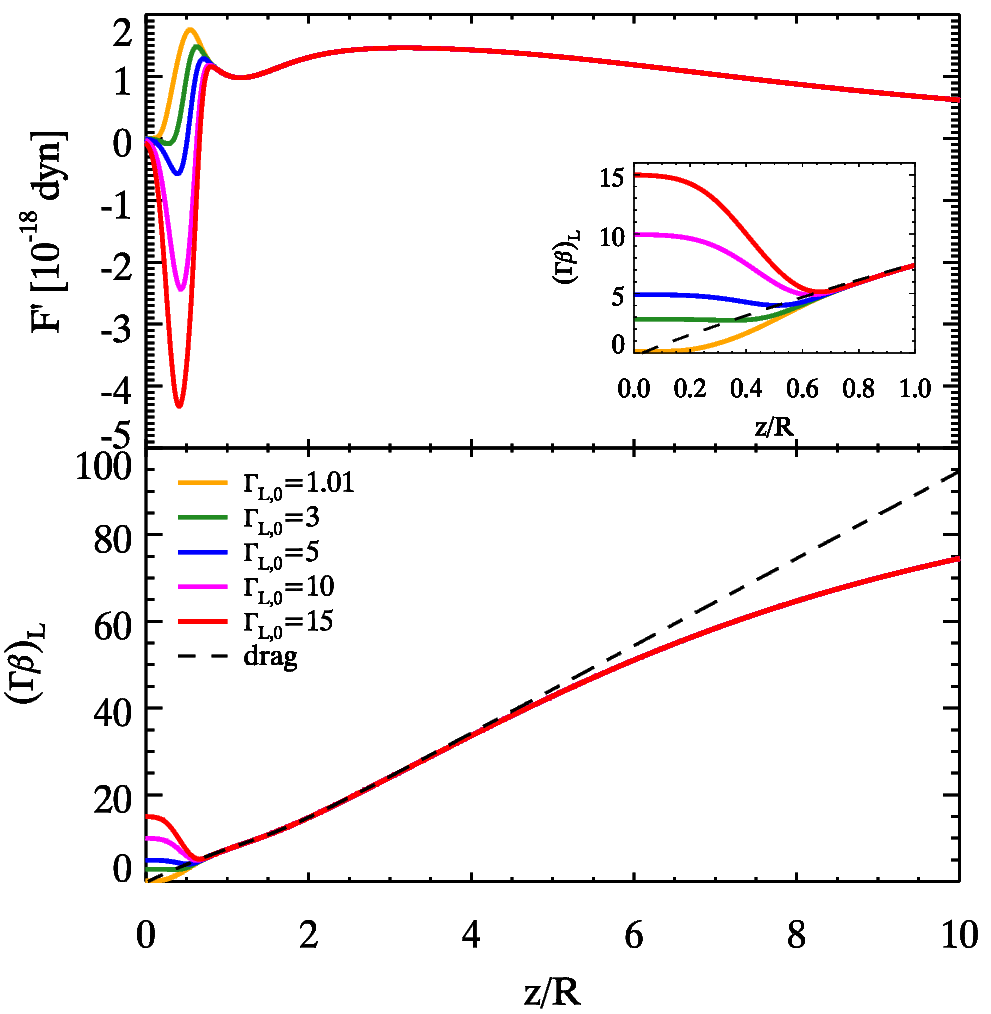}
\caption{
Radiative acceleration of the layer obtained by varying the initial layer Lorentz factor 
$\Gamma_{\rm L,0}=1.01$, $3$, $5$, $10$, $15$. 
The spine isotropic luminosity is $L_{\rm iso,S}=10^{45}$ erg/s and the 
leptons of the layer are hot ($\langle \gamma^2 _{\rm L}\rangle=10^6$) implying $k_L=10^{51}$ erg/s. The spine Lorentz factor is constant $\Gamma_{\rm S} = 15$.  
Top panel: radiative force $F^{\prime}$ in layer frame as a function of $z/R$. 
Bottom panel: layer Lorentz factor $\left(\Gamma \beta\right)_{\rm L}$ (solid lines) and the drag Lorentz factor 
$\left(\Gamma \beta\right)_{\rm L, drag}$ of the layer (dashed line) as a function of $z/R$; 
Inset: a zoomed view of the behaviour of the layer Lorentz factor for $0<z/R<1$.
}
\label{fig:nocoolseqgLhot}
\end{figure} 
%---------------------------------------

\vskip 0.2 cm
{\it Varying $k_{\rm L}$ ---}
Fig. \ref{fig:nocoolseqLg2} shows the effects of varying $k_{\rm L}$
%the product\footnote{
%We choose to consider the variation of the product 
%of $L_{\rm iso,S}$ and $\langle\gamma^2\rangle$ since both quantities 
%modify the conditions of the problem in the same way.}
%of the isotropic spine luminosity and the average random kinetic energy 
%$k=L_{\rm iso, S}\langle\gamma^2\rangle$ (which is our parameter for this case) 
for a constant spine Lorentz factor $\Gamma_S = 15$ and an initial bulk Lorentz 
factor of the layer $\Gamma_{\rm L,0} = 3$. 
At the base of the structure, i.e., for small values of $z$, the forces are negative for all the curves irrespective of the $k_{\rm L}$ value (the yellow curve has the greatest magnitude and hence is clearly visible below the zero force mark, whereas the other curves experience comparatively much smaller forces which are difficult to resolve on the scale of Figure~\ref{fig:nocoolseqLg2}). 
This is due to the fact that at the start ($z=0$) the layer particle sees a greater fraction of the radiation directed downward due to the entire spine--sheath structure located ahead. This decelerates the sheath thereby decreasing the $\left(\Gamma \beta\right)_{\rm L}$ as can be seen in the second panel, where the Lorentz factor profiles
are compared with the drag Lorentz factor. 
%The dashed line in the second panel denotes the drag Lorentz factor 
%i.e. the value of $\Gamma_{\rm L,drag}$ where the force,
%{\bf as measured in the comoving layer frame,} vanishes.
%{\bf Photons that hit the sheath ``effective particle'' with an angle, 
%as seen in the observer frame $K$, smaller than $1/\Gamma_{\rm L}$ 
%with respect to the direction of motion of the sheath
%are seen coming from upward in the sheath comoving frame.
%} 
%Notice that 
We start the simulations with a 
value of $\Gamma_{\rm L,0}=3$ which exceeds the drag Lorentz factor at that position.
As a result negative forces arise from the drag effects to reduce Lorentz factor at 
(or below) the drag level (refer to \S~\ref{sec:drag}).
We note that the force increases with the increasing values of 
$L_{\rm iso, S}\langle\gamma^2\rangle$ as seen from the 
force curves in the top panel of Fig. \ref{fig:nocoolseqLg2}.\\
For low spine luminosities or if the sheath plasma has small
internal energy (the case of a cold plasma with low value of 
$\langle\gamma^2_{\rm L}\rangle$) we find that the layer accelerates negligibly. 
Instead, significant acceleration is observed by increasing the spine luminosity 
or the mean squared energy of the particle in the layer.
In such cases, the bulk Lorentz factor profile $\left(\Gamma \beta\right)_{\rm L}$ 
manifests an initial decrease due to the initial negative force and then a subsequent increase. 
This increase, at the beginning, follows the drag Lorentz factor profile but it flattens afterwards 
depending upon $k_{\rm L}$:
%=L_{\rm iso, S}\cdot \langle\gamma^2\rangle$:
the greater the value of $k_{\rm L}$, the greater the final value of $\Gamma_{\rm L}$ 
and the longer the time for which the bulk Lorentz factor profile follows 
the $\Gamma_{\rm L,drag}$ profile.\\

The curves with the highest $k_{\rm L}$ ($k_{\rm L} = 10^{51}$ erg s$^{-1}$) display a peculiar force profile characterized by a double peak shape.
This behavior is due to a confluence of two effects.
Increasing the force increases $\Gamma_{\rm L}$ 
%which cannot exceed the drag limit at that respective position. 
and if it approaches and attempts to surpass the drag limit, 
the force rapidly decreases (refer to \S~\ref{sec:drag}). 
Thus, the drag effect is responsible for producing the rapid drop and consequently, 
the first peak in the force profile.
The second effect is due to the fact that 
when the layer effective particle surpasses the length--scale of the 
structure $R$, it receives most of the radiation produced by the entire spine length 
(pushing the effective particle along the positive $z$ direction), which lies behind the particle. This produces the second maximum of the force profile.
%
%switches off to prevent further acceleration and this results 
%in the drop/dip in the force curves. 
Note that the double peak is absent when the layer bulk Lorentz factor is nowhere near the drag limit.

% (specifically the peak followed by a rapid drop) in %
%the case of the curves for which $\left(\Gamma \beta\right)_{\rm L}$ values differ 
% significantly from the drag Lorentz factors. 

%
%also the reason behind why all the force 
%curves reach a maximum well beyond the $R=10^{16}$ cm mark, 
%which happens when the effective particle surpasses the length of the 
%structure $R$ and thereby receives radiation (almost all 
%of which tends to push the layer along positive $z$ direction) 
%not just from a section of the spine but from {\bf its entire length}. 
%The last argument can also explain how a single peak/maximum in the force curve arises 
%for cases where the layer Lorentz factor is nowhere near the drag limit.
% --------------------------------------
\begin{figure} 
\includegraphics[width=80mm]{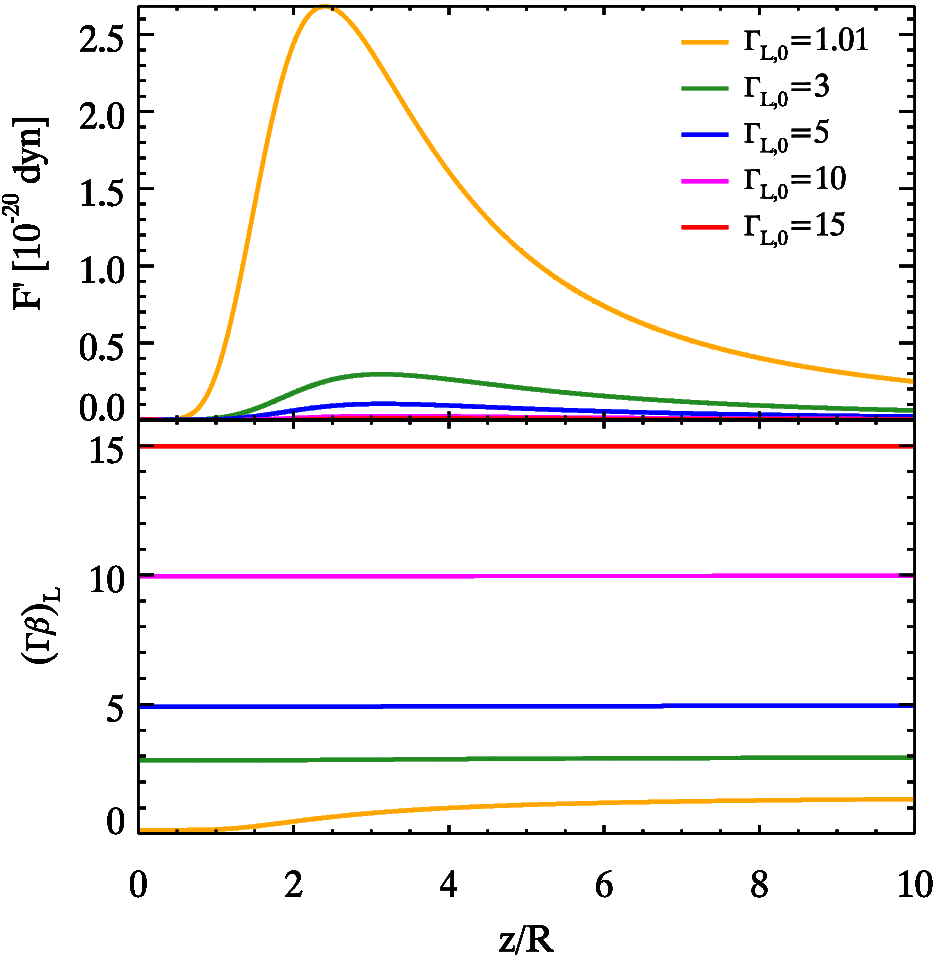}
\caption{
Radiative acceleration of the layer obtained by varying the initial layer Lorentz factor 
$\Gamma_{\rm L,0}=1.01$, $3$, $5$, $10$, $15$. 
The spine isotropic luminosity is  $L_{\rm iso,S}=10^{45}$ erg s$^{-1}$ and the 
leptons of the layer are cold ($\langle \gamma^2 _{\rm L} \rangle=1$). 
The spine Lorentz factor is constant $\Gamma_{\rm S} = 15$.  
Top panel: Radiative force $F^{\prime}$ in layer frame as a function of 
$z/R$. 
Bottom panel: Layer Lorentz factor $\left(\Gamma \beta\right)_{\rm L}$ (solid lines) as a function of $z/R$.}
\label{fig:nocoolseqgLcold}
\end{figure} 
% --------------------------------------
% --------------------------------------
\begin{figure} 
\includegraphics[width=80mm]{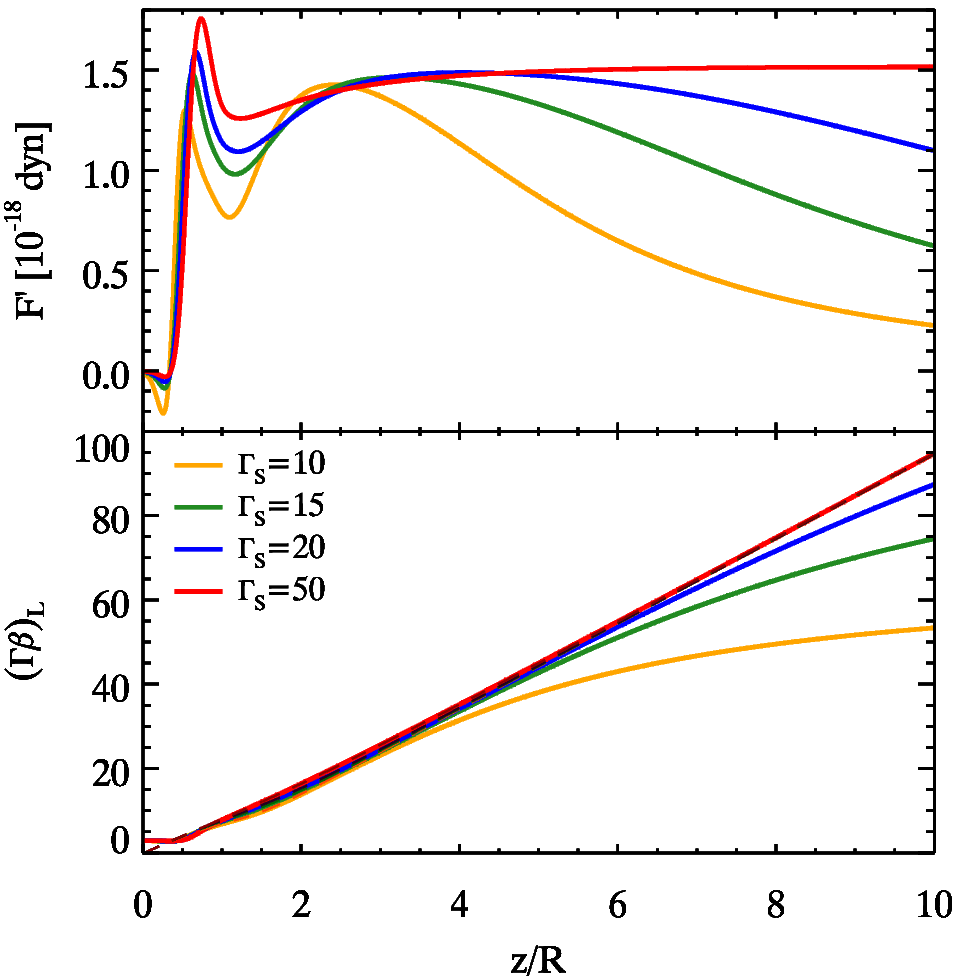} 
\caption{
Radiative acceleration of the layer due to variation of the spine bulk Lorentz factor $\Gamma_{\rm S}=10$, $15$, $20$, $50$. 
The spine isotropic luminosity is $L_{\rm iso,S}=10^{45}$ erg s$^{-1}$ and the leptons 
of the layer are hot ($\langle \gamma^2 _{\rm L}\rangle=10^6$). 
The initial layer Lorentz factor is constant $\Gamma_{\rm L,0} = 3$.
Top panel: radiative force $F^{\prime}$ in layer frame as a function of $z/R$.
Bottom panel: Layer Lorentz factor $\left(\Gamma \beta\right)_{\rm L}$ (solid lines) and 
the drag Lorentz factor $\left(\Gamma \beta\right)_{\rm L, drag}$ 
of the layer (dashed line) as a function of $z/R$.} 
%({\bf future}) inset: a zoomed view of the behaviour of the layer Lorentz factor for $0<z<R$.
\label{fig:nocoolseqgShot}
\end{figure} 
% --------------------------------------
% --------------------------------------
\begin{figure} 
\includegraphics[width=80mm]{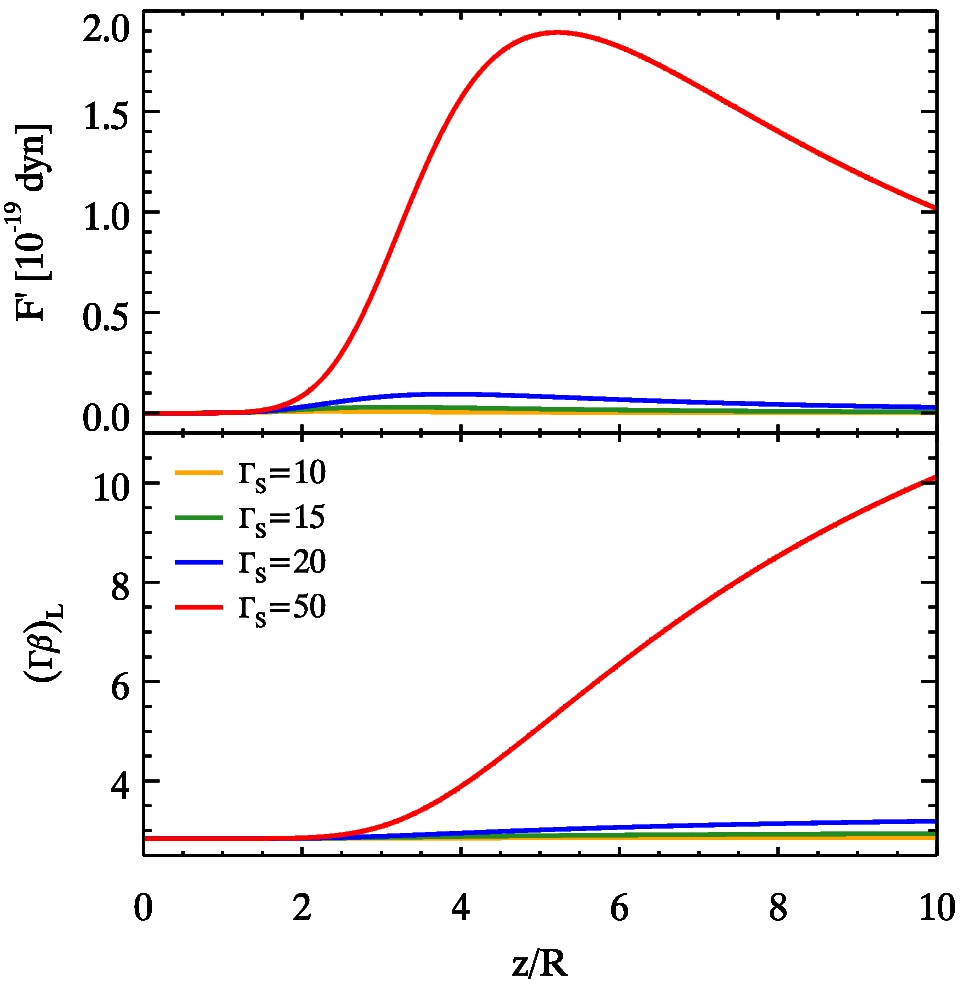}  
\caption{
Radiative acceleration of the layer obtained by varying the spine bulk Lorentz factor 
$\Gamma_{\rm S}=10$, $15$, $20$, $50$. 
The spine isotropic luminosity is $L_{\rm iso,S}=10^{45}$ erg s$^{-1}$ and the 
leptons of the layer are cold ($\langle \gamma^2 _{\rm L}\rangle=1$). 
The initial layer Lorentz factor is constant $\Gamma_{\rm L,0} = 3$.  
Top panel: radiative force $F^{\prime}$ in layer frame as a function of $z/R$. 
Bottom panel: layer Lorentz factor $\left(\Gamma \beta\right)_{\rm L}$ (solid lines) as a function of $z/R$.
}
\label{fig:nocoolseqgScold}
\end{figure} 
% --------------------------------------
%This explains the rise in the force curves and also of the 
%second peak of the blue and pink curves. 
%Thus, the greater the value of $L_{\rm iso, S}\langle\gamma^2\rangle$, 
%the greater are the forces and thus the sheath can attain higher values 
%of $\Gamma_{\rm L}$.\\
\vskip 0.2 cm
{\it Varying $\Gamma_{\rm L, 0}$ ---}
Fig.~\ref{fig:nocoolseqgLhot} shows the effects on radiative acceleration of 
the layer due to variation in the initial layer bulk Lorentz factor $\Gamma_{\rm L,0}$.
% (the parameter chosen for this case). 
The other quantities that remain fixed are the average internal energy content of the 
sheath leptons ($\langle\gamma^2_{\rm L}\rangle = 10^6$) and the isotropic spine luminosity 
$L_{\rm iso, S}=10^{45}$ erg s$^{-1}$. 
% In all variations explored,
In all cases, the force profile is 
characterized by the same features as observed in Fig.~\ref{fig:nocoolseqLg2}: 
initial negative force and double peak shape.
However due to the different values of $\Gamma_{\rm L,0}$, the force magnitudes for the various curves 
are initially different, with the curves traveling at larger $\Gamma_{\rm L,0}$ experiencing 
initially a larger force which decelerates them below the drag limit. 
%We note that the different curves cross the zero force line at different positions along the spine $z$ (upper panel of the zoomed view) which corresponds to crossing the dashed drag line in the bottom panel. 
We also note an interesting merging feature of both the $\Gamma_{\rm L}$ and the force curves. 
With identical forces and Lorentz factor $\Gamma_{\rm L}$, we expect and observe the 
trajectories of the curves to remain merged.
%{\bf Like in the previous case}, all the curves start with 
%sheath Lorentz factors ($\Gamma_{\rm L} > \Gamma_{\rm L,drag}$) 
%leading to negative forces which reduce the Lorentz factor of each of the curves as can be seen from the zoomed view in panel 2.  
This feature suggests that, in case of hot plasma and for $z>R$, the dynamical evolution of the layer 
does not depend on its initial Lorentz factor $\Gamma_{\rm L,0}$.

%Does this hint that Liso, Gspine and gamma square average are more important variables than initial GL0 value...
%We also note the emergence of the double force peaks for all the curves, which can be understood using the reasoning involving approaching the drag limit and receiving radiation from the entire spine as outlined in the previous paragraph. At large value of $z$, we expect from the bottom panel of figure 3 %\ref{fig5} \\
%that the Lorentz factors for the sheath would be large, a result similar to the result for the maximum product case (blue curve) in figure 2.\\ %\ref{?}.
Fig.~\ref{fig:nocoolseqgLcold} depicts the radiative evolution of a cold plasma 
($\langle\gamma^2 _{\rm L}\rangle = 1$) for an isotropic spine luminosity ($L_{\rm iso, S} = 10^{45}$ erg s$^{-1}$) 
and a constant spine Lorentz factor of $\Gamma_{\rm S} = 15$ . The value of $k_{\rm L}$ for 
Figs.~\ref{fig:nocoolseqgLhot} and \ref{fig:nocoolseqgLcold} differs by $10^6$ which is 
the average of the square of the leptonic Lorentz factor for the hotter plasma. 
We continue to vary $\Gamma_{\rm L,0}$ as our parameter 
and by comparing the two figures (Fig.~\ref{fig:nocoolseqgLcold} and Fig. \ref{fig:nocoolseqgLhot}) 
we note that the forces experienced by the colder leptons are smaller by two orders of magnitude. 
This strong reduction in the force results in negligible acceleration of the layer 
effective particle which maintains its initial Lorentz factor 
$\Gamma_{\rm L,0}$ except for the case $\Gamma_{\rm L,0}=1.01$, 
where there is a weak increase of the bulk Lorentz factor to $\Gamma_{\rm L,fin} = 1.67$.
%The reduction in force results first in reduced acceleration/deceleration which restricts the $\Gamma_{\rm L}$ from surpassing the $\Gamma_{L, Drag}$ value. 
We also note that as the forces involved are smaller than the previously considered cases, 
the drag force is not strong enough to create multiple force peaks. 
Thus the force profiles are characterized by single peaks which occur when the radiation 
from the entire spine irradiates the sheath particle.

%emerging at large $z$ values due to the bottle receiving photons from the entire spine. For the case of cold sheath leptons, overall we do not observe a significant change in the sheath Lorentz factors.\\
%These cases have higher aberration effects due to the large $\Gamma_L$ attained.\\

\vskip 0.2 cm
{\it Varying $\Gamma_{\rm S}$ ---}
In the final case for this subsection,
%  our parameter of interest is 
% the spine's Lorentz factor $\Gamma_{\rm S}$. 
we explore the effects of varying $\Gamma_{\rm S}$ on the force and 
$\left(\Gamma \beta\right)_{\rm L}$ curves for $\Gamma_{\rm L,0} = 3$. 
If the sheath plasma is hot, we observe double peaked curves 
as shown in the top panel of Fig.~\ref{fig:nocoolseqgShot}. %\ref{fig5}. 
We also note that a faster spine produces a greater force on the sheath, resulting in a faster sheath.
Fig.~\ref{fig:nocoolseqgScold} %\ref{fig6} 
shows the same physical quantities but for a cold sheath and we note that the 
forces involved are reduced by an order of magnitude. 
This does not result in a significant change of the layer Lorentz 
factors except for the case when $\Gamma_S=50$, 
%corresponding to the magenta curve, 
where a comparatively larger force leads to an accelerating sheath.
%
%% Does the magenta line cross the dotted magenta line multiple times???
%%
%Figures are double panel with profile of force and profile of $\Gamma_{\rm L}$.
%--- Sequence with varying $L_{\rm S} \langle \gamma^2\rangle$ 
%--- Sequence with varying $\Gamma_{\rm L,0}$ 
%--- explain in words what happens varying $\Gamma_{\rm S}$.
%--- Profile of $\gamma_{\rm drag}$ (overplotted)
%total: 4 figures.
% ---------------------------------------------
%\begin{figure} 
%\includegraphics[width=80mm]{New_0206_Liso_glsat2.png}	
%	\includegraphics[width=80mm]{Liso_avgsq_decay.png}	
%	\psfig{file=figure1cooling.png,width=19.2cm,height=16cm } 
%\caption{Radiative acceleration of the sheath obtained by varying the isotropic spine luminosity 
%$L_{\rm iso,L}$ from $10^{42}$ to $10^{47}$ erg s$^{-1}$ for sheath constituted by hot leptons 
%($\langle\gamma^2_{\rm L}\rangle = 10^6$), constant spine Lorentz factor of $\Gamma_{\rm S} = 15$ 
%and for an initial sheath Lorentz factor $\Gamma_{\rm L} = 3$. 
%The x--axis is common for all three panels and depicts the position $z$ normalized by 
%the vertical structure dimension $R$. 
%The top panel depicts the radiative force measured in the layer frame. 
%Middle panel: The internal energy of the sheath denoted by $\langle\gamma^2\rangle$. 
%Bottom panel: Lorentz factor of the sheath $\left(\Gamma \beta\right)_{\rm L}$ measured in the observer frame.}
%\label{fig:cool_Liso_gen}
%\end{figure}
% ---------------------------------------------
% ---------------------------------------------
\begin{figure}
\includegraphics[width=80mm]{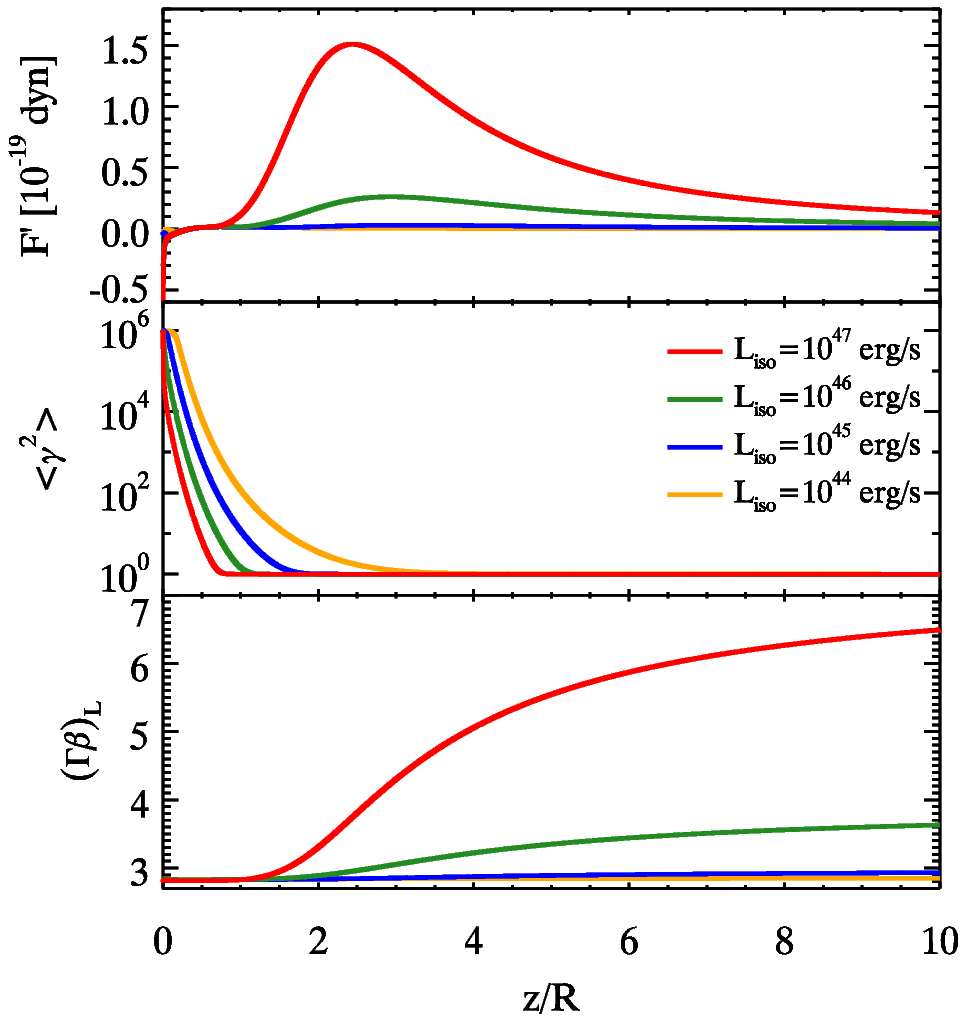}
\caption{
Radiative acceleration of the sheath obtained by varying the isotropic spine 
luminosity $L_{\rm iso,S}$ from $10^{44}$ to $10^{47}$ erg s$^{-1}$ for sheath constituted by 
hot leptons ($\langle\gamma_{\rm L}^2\rangle = 10^6$), constant spine Lorentz factor of $\Gamma_S = 15$ and for 
an initial sheath Lorentz factor $\Gamma_L = 3$. 
The x--axis depicts the position $z$ in units of the vertical structure dimension $R$. 
The top panel depicts the radiative force measured in the layer frame. 
Middle panel: the internal energy of the sheath denoted by $\langle\gamma_{\rm L}^2\rangle$. 
Bottom panel: Lorentz factor of the sheath $\left(\Gamma \beta\right)_{\rm L}$ 
measured in the observer frame.}
\label{fig:cool_Liso_glsat}
\end{figure}
% ---------------------------------------------

\subsection{Radiative acceleration of the layer: cooling scenario}
\label{cooling}
\subsubsection{Single injection}
\label{subsec:cool_nofeed}

In this subsection we explore a case where the particles of both the spine and the layer are energized only once before 
entering the active region and the particles in the layer can cool via radiative cooling.
This implies that  the layer leptonic energy distribution varies with time.
For simplicity, we will assume here that the spine bulk Lorentz factor $\Gamma_{\rm S}$ 
is constant, leaving the study of the possible change of $\Gamma_{\rm S}$ (caused by the interaction 
with the layer photons) to \S~\ref{subsec:feed}.

We start by solving the equation of motion for the layer and we show the spatial profiles 
for $\left(\Gamma \beta\right)_{\rm L}$, the evolution of the 
average internal energy of the layer 
$\langle\gamma^2_{\rm L}\rangle$ and the profile of the force as perceived 
by the layer projected over the $z$-axis $F^\prime$.

\begin{figure} 
\includegraphics[width=80mm]{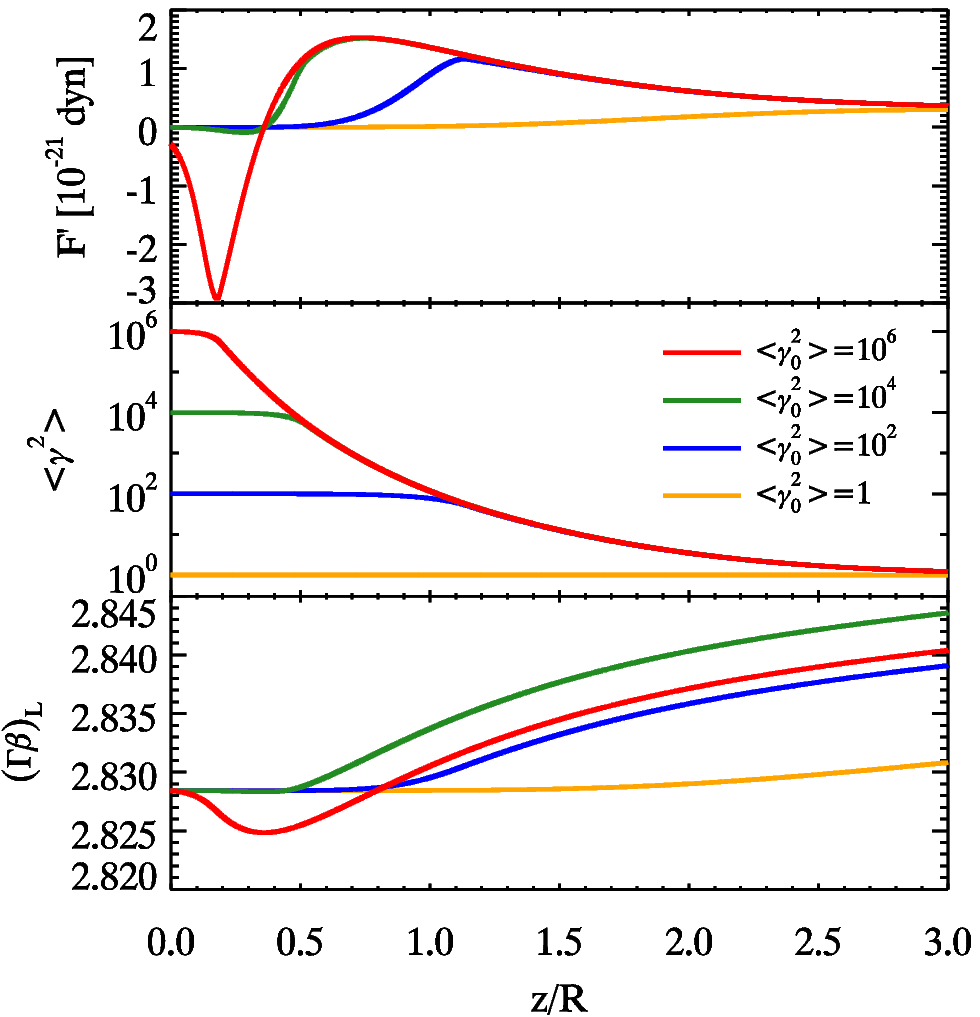}
\vskip 4pt
\caption{
Radiative acceleration of the sheath due to variation of the average internal energy content 
of the sheath $\langle\gamma^2_{\rm L}\rangle$ from $1$ to $10^6$ for a constant isotropic spine luminosity 
$L_{\rm iso,S}=10^{45}$ erg s$^{-1}$, a constant spine Lorentz factor $\Gamma_S = 15$ and for an initial 
sheath Lorentz factor $\Gamma_L = 3$. 
The x--axis depicts the position $z$ normalized by the 
vertical structure dimension $R$. 
Top panel depicts the radiative force measured in the layer frame. 
Middle panel: the internal energy of the sheath denoted by $\langle \gamma^2 \rangle$. 
Bottom panel: Lorentz factor of the sheath $\left(\Gamma \beta\right)_{\rm L}$ 
measured in the observer frame.}
\label{fig:cool_avgsq_gen}
\end{figure}

\vskip 0.2 cm
{\it Varying $L_{\rm iso,S}$ ---}
We begin by varying the intrinsic spine luminosity $L_{\rm iso,S}$, 
assuming that the population of the sheath is initially hot
($\langle\gamma^2_{\rm L}\rangle_{z=0} = 10^6$). 
The top panel of Fig.~\ref{fig:cool_Liso_glsat} depicts the force 
curves for different values of the spine luminosity.
For small values of $z$, the force profiles are similar to the profiles in Fig.~\ref{fig:nocoolseqLg2}. 
All the curves show an initial negative force with magnitudes proportional 
to the spine luminosity (refer to \S~\ref{subsec:nocool_nofeed}).
%One feature that the force curves in fig. \ref{fig:cool_Liso_gen}
%share with the curves in fig \ref{fig:nocoolseqLg2} 
%are the initial negative forces with magnitudes proportional to the spine luminosity. 
The middle panel shows the variation of $\langle\gamma^2_{\rm L}\rangle$ and we note the difference 
in the cooling rates for the various curves, which arises because sheaths with more 
luminous spines cool faster.
% The sheath with the greatest intrinsic spine luminosity cools the fastest (yellow curve) 
% and hence the greater the luminosity of the spine the faster is the radiative cooling of the sheath. 
Note that the various curves in the second panel eventually merge due to continuous cooling. 
The third panel shows the variation of $\left(\Gamma \beta\right)_{\rm L}$. 
For $z<R$, all the layer curves show no acceleration. On the contrary, from the force curves one expects the layer to be decelerated due to the initial negative forces but these decelerations are small and hence difficult to resolve (and thus, see) in Fig.~\ref{fig:cool_Liso_glsat}. However, the reader can observe the resolved deceleration for the blue curve in Fig.~\ref{fig:cool_Liso_glsat} through the red curve in the bottom panel of Fig.~\ref{fig:cool_avgsq_gen} because the product $k_L=10^{51}$ erg/s is identical for these curves.
%These small decelerations though can be observed in  where the red curve is identical to . exposed to higher luminosity experience greater decelerations due to more negative 
%initial forces. 
When the effective sheath particle overtakes the scale--length $R$, the spine which lies behind it 
irradiates the particle from the rear. However, at these large values of $z$ the sheath plasma is cold and only extremely large spine luminosities 
(e.g. $L_{\rm iso,S} \geq 10^{45}$ erg s$^{-1}$) are capable of significantly accelerating the layer particle and hence the layer, as shown in Fig. \ref{fig:cool_Liso_glsat}.

% --------------------------------------------------

% --------------------------------------------------
\begin{figure} 
\includegraphics[width=80mm]{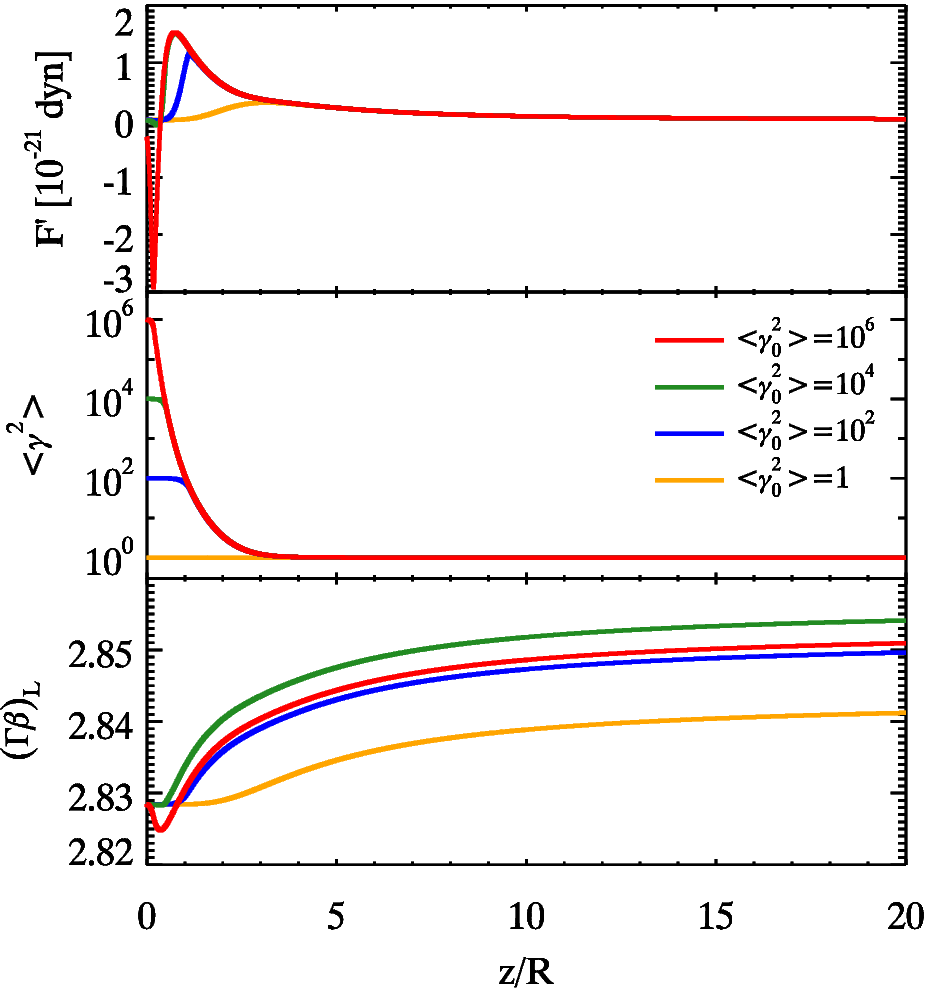}
\caption{
Same as Fig. \ref{fig:cool_avgsq_gen}, but zooming out to longer $z/R$ to 
see the final values of $\left(\Gamma \beta\right)_{\rm L}$.
}
%Radiative acceleration of the sheath obtained by varying the average internal 
%energy content of the sheath $\langle\gamma^2_{\rm L}\rangle$ from $1$ to $10^6$ for a constant 
%isotropic spine luminosity $L_{\rm iso,L}=10^{45}$ erg s$^{-1}$, a constant spine Lorentz factor 
%$\Gamma_{\rm S} = 15$ and for an initial sheath Lorentz factor $\Gamma_{\rm L} = 3$. 
%The x--axis depicts the position $z$ normalized by 
%the vertical structure dimension $R$. 
%The top panel shows the radiative force measured in the layer frame. 
%Middle panel: the internal energy of the sheath denoted by $\langle\gamma^2\rangle$. 
%Bottom panel: Lorentz factor of the sheath $\left(\Gamma \beta\right)_{\rm L}$ measured 
%in the observer frame.}
\label{fig:cool_avgsq_glsat}
\end{figure}

% --------------------------------------------------
%\begin{figure} 
%	\vskip -0.6 cm
%	\hskip -1.3 cm
%	% \psfig{file=1028r.ps,width=19.2cm,height=16cm } 
%	\vskip -7.3 cm
%	\caption{Vary Lspine Gamma Saturation}
%	\label{fig8}
%\end{figure}
%\begin{figure} 
%	\vskip -0.6 cm
%	\hskip -1.3 cm
%	% \psfig{file=1028r.ps,width=19.2cm,height=16cm } 
%	\vskip -7.3 cm
%	\caption{Vary Initial av gamma squared}
%	\label{fig8}
%\end{figure}
%
%
%\begin{figure} 
%	\vskip -0.6 cm
%	\hskip -1.3 cm
%	% \psfig{file=1028r.ps,width=19.2cm,height=16cm } 
%	\vskip -7.3 cm
%	\caption{Cooling Cold}
%	\label{fig10}
%\end{figure}

\vskip 0.2 cm
{\it 
Varying $\langle\gamma^2_{\rm L}\rangle_{z=0}$ ---}
We explore how the sheath evolves for different initial $\langle\gamma^2_{\rm L}\rangle_{z=0}$ (in short, $\langle\gamma^2_{0}\rangle$) values 
under the influence of radiative cooling with an initial sheath Lorentz factor $\Gamma_{\rm L,0} = 3$ 
and a constant spine luminosity $L_{\rm iso,S}=10^{45}$ erg s$^{-1}$. 
Fig. \ref{fig:cool_avgsq_gen} confirms that the force profile is characterized by negative values
during the initial stages, with the hottest sheaths experiencing the greatest force magnitudes.
% The bottom panel of figure \ref{fig:cool_avgsq_gen} informs us that the hotter plasma 
% is decelerated more due to a larger force in comparison with the other curves which 
% are comparatively cooler thereby the yellow and magenta curves show noticeably reduced velocity. 
The middle panel depicts the variation in $\langle\gamma^2_{\rm L}\rangle$ due to radiative cooling. 
The curves show an initial flat evolution and then a decreasing trend in such a manner that all the 
curves merge, irrespective of their $\langle\gamma^2_{0}\rangle$ values.
This behavior can be understood from Fig.~\ref{fig:average} or from particle energy distribution $N(\gamma)$ of the layer 
(see Eq. \ref{eq:ngamma}): if the cooling energy $\gamma_c$ is greater than the spectral 
break energy $\gamma_b$ (i.e., $\gamma_{\rm c}>\gamma_{\rm b}$), then $\g2av \sim \gamma_{\rm b}$ 
which implies that $\g2av$ is almost constant (this corresponds to the initially 
flat evolutionary phase seen in the middle panel of Fig. \ref{fig:cool_avgsq_gen}). 
However, when $\gamma_{\rm c}$ crosses $\gamma_{\rm b}$, $\g2av$ starts to decrease from its initial value 
and becomes comparable to $\gamma_{\rm c}$.
% as the bottle propagates along the z direction with cooling being the fastest for the sheath plasma 
% having the hottest leptons. It also depicts a merger of $<\gamma^2>$ values as the various plasmas 
% cool, with the corresponding force curves in the upper panel approaching one another. 
Note that even though the force curves have merged (simultaneously with the merging of 
$\g2av$ curves), the $\left(\Gamma \beta\right)_{\rm L}$ 
curves remain segregated due to the different initial decelerations 
resulting from the different initial behavior of $\g2av$. 
This leads to the interesting result -- the bulk Lorentz factor at saturation is maximum for an 
intermediate value of the initial $\langle\gamma_{\rm L}^2\rangle$ ($\langle\gamma_{\rm L}^2\rangle=10^4$), 
instead of the curve with the initially hottest leptons ($\langle\gamma_{\rm L}^2\rangle=10^6$), 
albeit by a small amount (see Fig.~\ref{fig:cool_avgsq_glsat}).

As a whole, we conclude from this section that the radiative cooling 
process causes the layer to lose internal energy rapidly 
and so it quenches the process of radiative acceleration.
Except for cases with very high spine luminosity, we can affirm that
the final bulk Lorentz factor of the layer $\Gamma_{\rm L}$ does not change
significantly from its initial value $\Gamma_{\rm L,0}$ (see Fig.~\ref{fig:cool_avgsq_glsat}). 
This result is similar to those  obtained by studying the radiative acceleration of a 
cold plasma in the no cooling scenario (cfr. \S \ref{subsec:nocool_nofeed}).
% leading to Lorentz factors saturating at $\Gamma_{\rm L} \sim 3$.
% we fix something ($\Gamma_{\rm L, 0}$ ??)
% --- 3 panels: 1: the force; 2: $\langle \gamma^2\rangle$, 3: $\Gamma_{\rm L}$
% --- Additional figure: Profile of the force explained: one additional reason to have a double peak
% We have to consider separately $L_{\rm S}$ and $\langle \gamma^2\rangle$
% Saturation of $\Gamma_{\rm L}$ 
% Sequence using different $L_{\rm S}$
% Sequence using different {\it initial} $\langle \gamma^2 \rangle$
% Sequence using different initial $\Gamma_{\rm L}$ ???

\subsubsection{Continuous injection}
\label{subsec:continj}

This scenario explores the situation where we have continuous energy injection inside the 
sheath region via, e.g., a standing shock, namely  when the cooling is balanced by injection of
fresh energetic particles that energize the plasma and the electron energy distribution is assumed to stay constant between $z=0$ and $z=10^{16}$ cm. 
There is no energy injection outside the active region. 
We assume that the injection rate within the active region is 
such that its effect is equivalent to making cooling ineffective. 
We still neglect here the radiative effects on the spine due to the layer and hence the bulk 
Lorentz factor of the spine $\Gamma_{\rm S}$ is assumed to be constant.\\

Differently from most of the other cases presented in this work, for this case
we assume that the initial bulk Lorentz factor of the layer is close to unity.
In fact we intent to explore if the layer can be accelerated radiatively and, achieve a bulk Lorentz factor $\Gamma_{\rm L} \sim 3$ as required by observations.
%In this section we neglect the feedback on the dynamics of the spine,
%analyzed later, to compare the results obtained here to those obtained by  including the feedback mechanism.

% --------------------------------------------------
\begin{figure} 
\includegraphics[width=80mm]{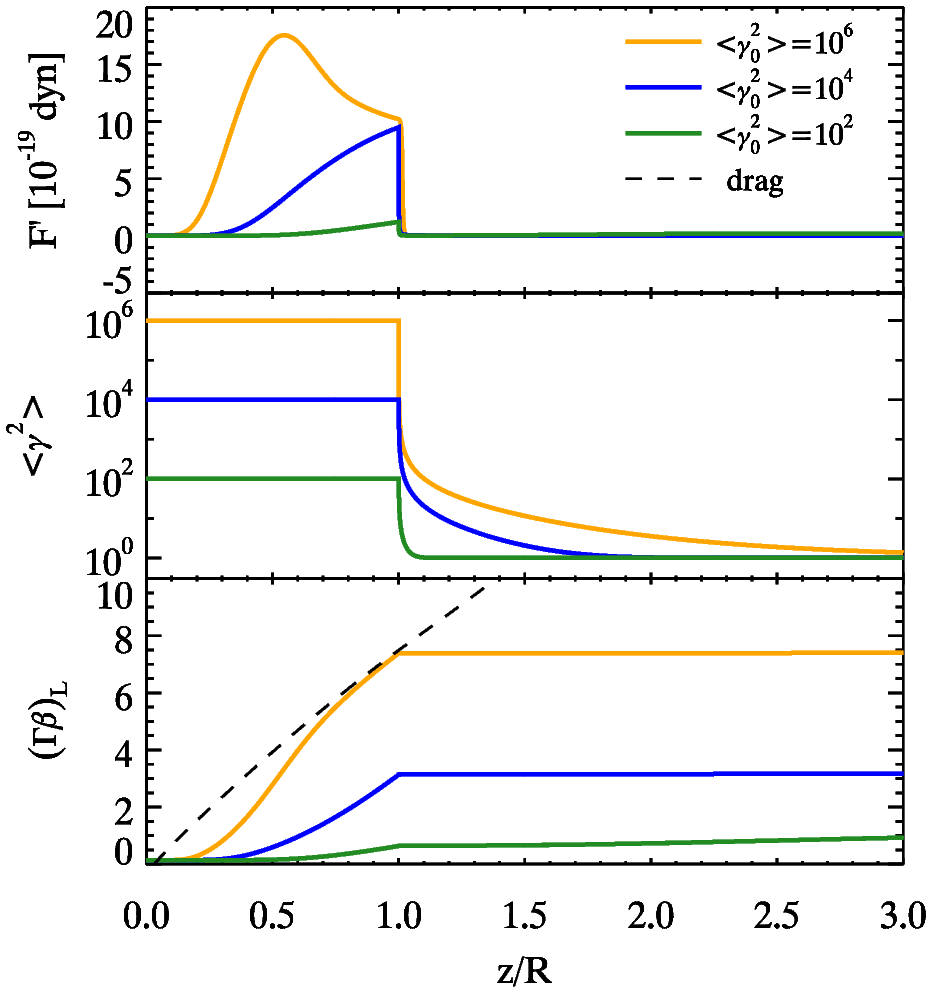}
\caption{ 
Radiative acceleration of the sheath obtained by varying the initial average internal energy content of 
the sheath $\langle\gamma^2_{\rm 0}\rangle$ from $10^2$ to $10^6$ for a constant isotropic spine luminosity 
$L_{\rm iso,S}=10^{45}$ erg s$^{-1}$, a constant spine Lorentz factor $\Gamma_{\rm S} = 15$ and for an 
initial sheath Lorentz factor $\Gamma_{\rm L, 0} = 1.01$. 
In this scenario, the plasma cools radiatively only for $z/R>1$; for $0<z/R<1$ the plasma is continuously 
energized and maintains the average internal energy (thereby making cooling ineffective). 
The top panel depicts the radiative force measured in the layer frame. 
Middle panel: The variation in the internal energy of the sheath denoted by 
$\langle \gamma^2_{\rm L} \rangle$ 
as a function of the vertical position $z$. 
Bottom panel: $\left(\Gamma \beta\right)_{\rm L}$ profile  of the sheath as 
measured in the observer frame.}
\label{fig:contshocks}
\end{figure}
% --------------------------------------------------

Fig. \ref{fig:contshocks} depicts the evolution of the force, $\langle \gamma^2 \rangle$ and 
$\left(\Gamma \beta\right)_{\rm L}$ curves. 
The continuous shocking re-energizes the particles within the standing shock region, 
% i.e., for $0\leq z\leq R$, rendering the cooling ineffective and 
and it is responsible for large forces proportional 
to the plasma's internal energy $\langle \gamma^2_{\rm L} \rangle$ content. 
Because there is continuous energy injection for $z\leq 10^{16}$ cm, we expect the physical 
quantities in this region to evolve in a fashion similar to the same quantities 
in \S ~\ref{subsec:nocool_nofeed}.
Indeed, the evolution of the force curves in the standing shock region as depicted in 
Fig. \ref{fig:contshocks} is  very similar to Fig.~\ref{fig:nocoolseqLg2} 
(the yellow curves are almost identical for $z<R$). 
The drop in the force is due to the bulk Lorentz factor of the layer approaching 
$\Gamma_{\rm L, drag}$ as explained earlier in \S ~\ref{subsec:nocool_nofeed}. 
All the three curves experience positive forces, leading to an accelerated layer as 
demonstrated by the $\left(\Gamma \beta\right)_{\rm L}$ plot in the bottom panel of 
Fig.~\ref{fig:contshocks}. 
The cooling becomes effective for $z>10^{16}$ cm ($z/R>1$), 
resulting in a rapid decrease in the values 
of $\langle \gamma^2_{\rm L} \rangle$ for the various curves (middle panel of Fig.~\ref{fig:contshocks}). 
A decrease in $\langle \gamma^2_{\rm L} \rangle$ correspondingly produces a rapid decrease in the force 
values also (top panel of Fig. \ref{fig:contshocks}). 
The role of radiative cooling in quenching the radiative force has been explored in 
detail in \S ~\ref{subsec:cool_nofeed}. 
As the force values for $z>10^{16}$ cm are quite small, we note that the layer Lorentz 
factors saturate close to the values attained at $z\sim R=10^{16}$ cm. 

Fig. \ref{fig:contshocks} shows that if the layer is kept hot by a mechanism replenishing
its energy losses, it can indeed be accelerated to ``interesting" Lorentz factors 
(i.e. $\Gamma_{\rm L}\sim 3$).
Of course, the hotter the layer, the stronger the Compton rocket effect and the larger
the final Lorentz factor.
We will see in the next subsection if this remains true when considering the feedback
on the spine.

% \textbf{\textit{We note in this case the saturation values of $\Gamma_{\rm L}$ is about 2.74 
% (blue curve in Fig.~\ref{fig:contshocks}) -- a value which is small enough to prevent strong 
% de-beaming and large enough to suppress the $\gamma \gamma$ absorption process 
% (Ghisellini, Tavecchio \& Chiaberge 2005).---AC--- so the idea is that we are getting 
% the layer to have a Lorentz factor $\sim$ 3 which is what we were hoping for..?}}\\
% comparable to $\Gamma_L \sim 3$ .

\subsection{Spine--layer feedback}\label{subsec:feed}

% --------------------------------------
\begin{figure} 
\includegraphics[width=80mm]{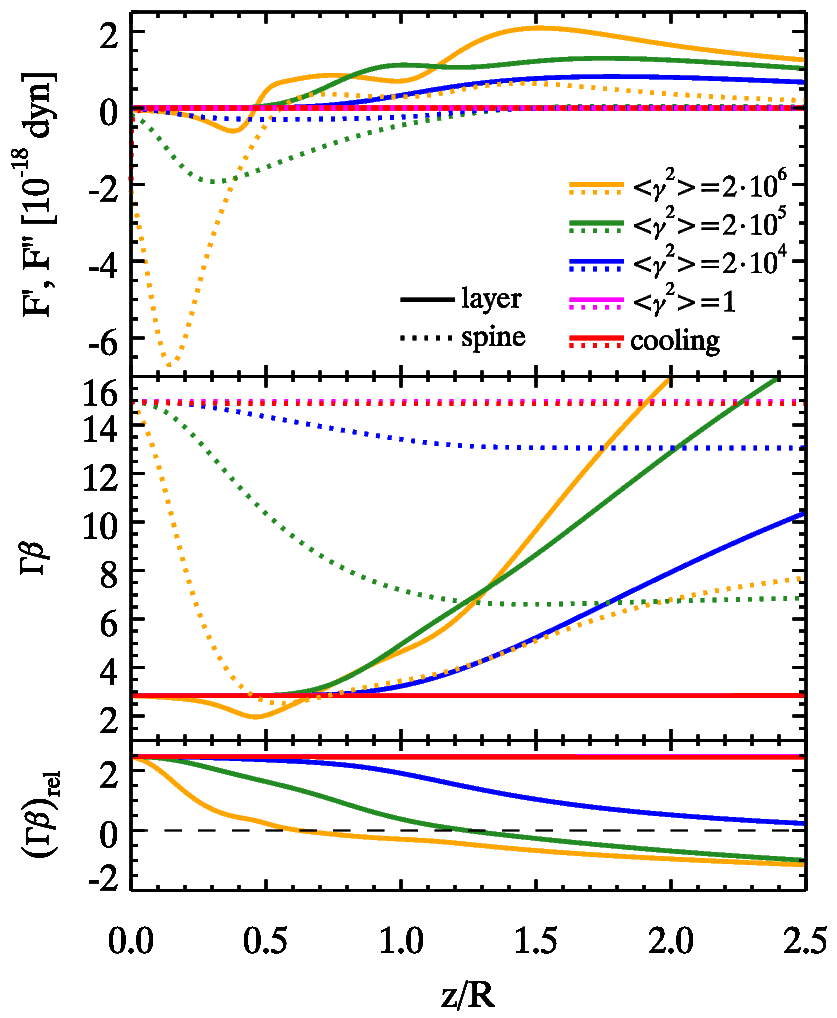}
\caption{
Radiative acceleration of spine and layer: the feedback. 
Comparison by varying $\langle \gamma^2 \rangle=2 \cdot 10^6$, 
$2 \cdot 10^5$, $2 \cdot 10^4$, $1$ of both spine and layer in no cooling 
case and cooling case.
Parameters used: $\Gamma_{\rm S, 0}=15$, $\Gamma_{\rm L,0}=3$ 
and $L^{\prime\prime}_{\rm S}=10\cdot L^{\prime}_{\rm L}$.
Top panel: radiative force measured in spine (layer) frame in dotted 
(solid) lines as a function of $z/R$.
Middle panel: Profile of $\left(\Gamma \beta\right)$ of the spine (layer) in 
dotted (solid) lines. 
Only for $\langle \gamma^2 \rangle= 2\cdot10^6$ 
with no cooling, in the final stages the layer can travel faster than the spine. 
Bottom panel: The relative velocity profile of the spine with respect to the layer 
expressed in terms of $\left(\Gamma \beta\right)_{\rm rel}.$}
\label{fig:feedback}
\end{figure} 
% --------------------------------------

In \S \ref{subsec:feedb}, we described the feedback mechanism produced by relative interaction 
between the photons emitted by the layer and the spine particles and vice--versa, i.e., 
between spine photons and layer particles. 
In this section we aim to study the spine--layer feedback mechanism and how the mechanism 
modifies the Lorentz factor profiles for both the spine and the layer. 
First we will compute both the spine and the layer bulk Lorentz factor profiles 
[$\Gamma_{\rm S}(z)$ and $\Gamma_{\rm L}(z)$] in a self--consistent manner.
Second, we will study two cases with and without the presence of radiative cooling: 
in the first case the particle energy distribution is fixed for both the spine and the 
layer, which is equivalent to the no cooling scenario (however here feedback is active). 
In the other case, both the spine and layer particles cool radiatively (with feedback).

The spatial profiles of the rest frame forces and of the 
bulk Lorentz factors for spine and layer have been computed by
numerically solving Eqs. \ref{eq:feedback}. 
The results for the two cases (with cooling and without cooling), mentioned earlier
are shown in Fig. \ref{fig:feedback} and are be summarized below:

% --------------------------------------
\begin{enumerate}

\item As expected, in all the explored cases the spine is decelerated 
by the Compton interaction with the layer photons. The force 
experienced by the spine is always negative (the only exception is the case
with $\g2av \sim 10^6$, where in the last stages the spine force turns positive);

\item consistent with the results of previous sections, 
the layer is initially decelerated and later accelerated in all cases; 

\item the acceleration is stronger for greater values of $\g2av$ and is
negligible for cold plasmas; 

\item the cooling case reproduces the results of a cold spine 
and a cold layer, i.e., no significant change of bulk Lorentz factor;

\item for high values of $\g2av \sim 10^6$ the effect of deceleration of the
spine and acceleration of the layer is so strong that the sheath can eventually
travel faster than the spine. 
This occurs because the cooling has been switched off.
We believe that this is rather unrealistic: to maintain such an energetic particle distribution a huge amount of energy must 
be supplied to the plasma.
\end{enumerate}
% --------------------------------------
\subsubsection{Continuous injection with feedback}\label{subsec:feedback:continj}
\begin{figure} 
	\includegraphics[width=80mm]{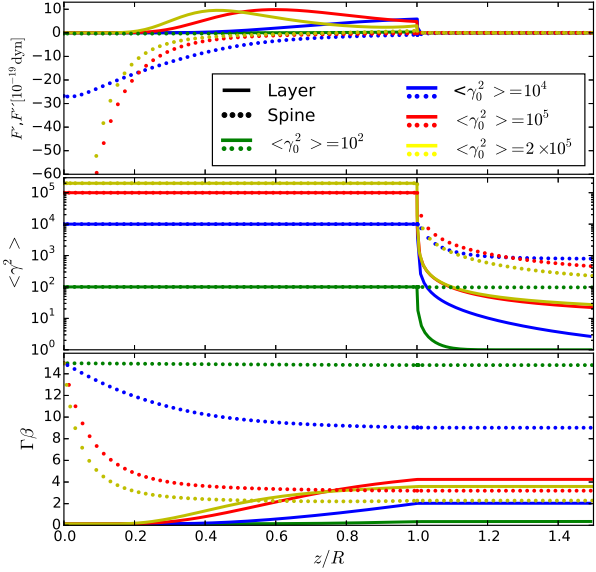}
	\caption{
		Radiative acceleration of spine and layer: the feedback with continuous injection. In this scenario, the plasma radiatively cools for $\rm z/R> 1$ whereas for $0<z/R<1$ it is continuously energized thereby nullifying the radiative cooling (thus maintaining its internal energy content). The colored curves are obtained by varying $\langle \gamma^2 \rangle=2\times 10^5$, $10^5$, $10^4$ and $10^2$ for both the spine and the layer.
		Parameters used: $\Gamma_{\rm S, 0}=15$, $\Gamma_{\rm L,0}=1.01$ 
		and $L^{\prime\prime}_{\rm S}=10\cdot L^{\prime}_{\rm L}$.
		Top panel: radiative force measured in spine (layer) frame in dotted 
		(solid) lines as a function of $z/R$.
		Middle panel: The variation in the internal energy of content of the spine and the layer denoted by $\langle \gamma^2_{\rm L} \rangle$ as a function of the vertical position $z$. 
		Bottom panel: $\left(\Gamma \beta\right)$ profile measured in the observer frame.}
	\label{fig:feedback_continject}
\end{figure} 
% --------------------------------------
%\begin{figure} 
%	\includegraphics[width=80mm]{lessdata_dots_intermediate_feedback.eps}
%	\caption{
%		LESS DATA DOTS INTERMEDIATE FEEDBACK EPS
%	}
%	\label{fig:feedback_continject}
%\end{figure} 
%%%%
In this scenario we study the kinematic evolution of the plasma due to continuous injection (see \S \ref{subsec:continj}) while taking into account the effects of spine--layer feedback. The initial internal energy content of the spine and the layer plasmas ($\g2av_S$ and $\g2av_L$ respectively) is maintained by continuous injection of particles within the region $0 \leq z/R < 1$ thereby rendering radiative cooling ineffective. However, in the absence of particle injection, i.e., for $z/R>1$ radiative cooling becomes important and impacts the evolution of the plasma by altering its internal energy content (see \S~
\ref{sec:cool}). This scenario differs from \S~\ref{subsec:continj} because here we account for the radiative interplay between the spine and the layer  and, as a result, the bulk Lorentz factor and the average internal energy content of the spine are no longer constants.\\
In order to study if the layer can accelerate from almost %\textasciiacute 
\textquoteleft at rest\textquoteright \space
%\textasciigrave 
situation and achieve bulk motions required for explaining observations, we begin with the layer moving at a small initial bulk Lorentz factor $\Gamma_{\rm L,0} = 1.01$. The three panels, from top to bottom in Fig.~\ref{fig:feedback_continject} show how the forces, $\g2av$, and, the velocity profile $\Gamma \beta$ for the spine and the layer evolve as a function of the normalized position $z$. We summarize the results of Fig.~\ref{fig:feedback_continject} below:
% --------------------------------------
\begin{enumerate}
	
	\item For the cases with different initial $\langle \gamma^2_0 \rangle$ values, the spine predominantly experiences a negative force with a magnitude proportional to the spine's internal energy content $\langle \gamma^2_S \rangle$. The forces are quite small for $\g2av=10^2$ and as a result negligible deceleration occurs in this case. The forces are significantly larger for cases with $\g2av \geq 10^5$ which result in the decelerating spine traveling slower than the layer. With the layer now moving faster than the spine, this \textquoteleft role reversal\textquoteright \space causes the spine to experience a positive and accelerating force. However, the magnitude of this force is small due to the reduction in the relative bulk Lorentz factor between the spine and the layer;
	\item the layer is first decelerated and then accelerated (due to the system's geometry as explained in detail in \S~\ref{subsec:nocool_nofeed}) as the top panel of Fig.~\ref{fig:feedback_continject} shows. However, for cases where $\g2av \geq 10^5$ the top panel depicts formation of a peak, followed by a decline in the layer force curves within the region $0 \leq z/R < 1$. The peak's location coincides with the location where the decelerating spine slows down to speeds comparable to those of the layer (see bottom panel) - leading to reduced relative motion which suppresses Compton rocket acceleration. As a result, the force experienced by the layer peaks and declines;
	\item the net acceleration of the layer does not generally correlate with the average internal energy content of the layer.
	%the net acceleration of the layer is in general not greater for hotter layers. 
	This is evident from the hotter plasma (yellow curve - $\langle \gamma^2_0 \rangle= 2\times10^5$) saturating at a lower speed than a cooler plasma (the red curve - $\langle \gamma^2_0 \rangle= 10^5$) as depicted in the bottom panel of Fig.~\ref{fig:feedback_continject}. This phenomenon can again be linked with a hotter spine being more quickly decelerated than a comparatively cooler spine. For sufficiently hot plasmas, the decelerating/slowing spine can end up traveling at comparable or slower speeds than the layer which leads to lesser relative motion and, thereby, suppression of the Compton rocket mechanism;
	\item in the $\g2av = 10^4$ case (blue curve) the layer accelerates until $z/R = 1$, after which particle injection ceases, and radiative cooling rapidly decreases the layer's internal energy content. As a result the gradually increasing radiative force vanishes and the acceleration process is quenched. Rapid radiative cooling also occurs for plasmas with $\g2av > 10^4$ (as shown by the red and yellow curves), and cools these plasmas faster than the $\g2av = 10^4$ case (blue curve) as shown in the middle panel of Fig.~\ref{fig:feedback_continject}. Again, this leads to the sudden drop in the force curves and quenches any further acceleration. The velocity saturation and force quenching phenomena observed here are identical to those observed in \S~\ref{subsec:continj}.
	
\end{enumerate}
As shown in the bottom panel of Fig.~\ref{fig:feedback_continject}, by replenishing the energy lost radiatively by injecting particles continuously the layer can attain and maintain \textquotedblleft interesting\textquotedblright \space Lorentz factors ($\Gamma_{\rm L} \sim 2 - 4.5$). Comparison between the results obtained here with those of \S~\ref{subsec:continj} shows that radiative feedback in hotter plasmas (both the spine and the layer are hot) reduces the relative bulk motion between the two components of the jet (i.e., the spine and the layer). This reduction in $\Gamma_{\rm rel}$ reduces the amount of boosted radiation thereby suppressing the Compton rocket effect. As a result, the layer does not accelerate and attain Lorentz factors beyond $\sim 4.5$. Thus in this particular scenario, acceleration/deceleration due to radiative feedback is responsible for regulating the bulk Lorentz factors of both the spine and the layer.
\subsection{Spine--layer feedback in e$^+$e$^-$ pair loaded plasmas}
\label{subsec:pairsfeed}

In this subsection we explore the impact of electron--positron pair loading on the radiative 
acceleration of the spine and the layer. 
While accounting for radiative feedback between the spine and the layer we explore the effects 
that different amounts of pair concentrations can have on the spine--layer system. 
The main effect of pairs will be to make the plasma ``lighter", in the sense that the
radiative force will act on an increased number of leptons, while the inertia
is still dominated by the same number of protons (except for the pair--dominated cases).
Therefore the acceleration or the deceleration will be stronger. 
The amount of pairs in the plasma can be characterized by the lepton to proton ratio $f$ (as defined in \S \ref{subsec:eqmotion}) as follows:
% --------------------------------------
\begin{itemize}

\item Pair--free (PF) plasma ($f=1$): there is one proton for every electron and no pairs are present.

\item Pair enriched (PE) plasma ($f\sim20$): this plasma is characterized by the presence of 
several electron--positron pairs. The value $f=20$ is about the maximum allowed 
from considerations about the total power of relativistic jets (Ghisellini \& Tavecchio 2010).

\item Pair dominated (PD) plasma ($f \rightarrow \infty$): in this case leptons dominate 
the kinematics of the plasma. We show this case for illustration, even if it may not be realistic
for AGN jets.

\end{itemize}
% --------------------------------------
We will first study the case without feedback and with the spine moving with a constant
bulk Lorenz factor $\Gamma_{\rm S}=15$. 
Then we will study the feedback case, assuming that both the layer and the 
spine have the same number $f$ of pairs.
Within the case without feedback, we will study the extreme cases of a cold plasma
($\langle \gamma^2 \rangle = 1$, Fig.~\ref{fig:pairscold})
and a hot plasma ($\langle \gamma^2 \rangle = 10^6$, Fig. \ref{fig:pairshot}).
Finally, we study the feedback case (Fig. \ref{fig:feed_pairs}).
The results of these cases are summarized as follows:
% --------------------------------------

\begin{enumerate}
\item {\it Cold plasma -- No feedback ---} 
Fig.~\ref{fig:pairscold} depicts the force experienced by the layer due to radiation 
from the spine in the upper panel. 
The lower panel depicts the radiative acceleration of the 
layer by plotting $\left(\Gamma \beta\right)_{\rm L}$ as a function of position $z/R$. 
We note that the force magnitudes are extremely small $\sim 10^{-21}$ dyn in comparison with 
forces observed in the earlier sections. 
The pair--free plasma case is identical to $\langle \gamma^2 \rangle=1$ in \S~\ref{subsec:feed}. 
%{\bf TO BE REVISED IF WE DROP THE COLD CASE IN \S~\ref{subsec:feed}.
%HERE THE SPINE CAN BE HOT AND THE LAYER COLD - 
%\\--AC-- \\
%I have deleted any mention of the cold case in the summary of results in sec 3.3 - there is a mention of the cold case in the figure itself and it would be hard to re-do the figure as Francesco has the necessary data and access to IDL. So let us keep it in the figure\\
% --AC--}
Even though the forces are small, the increased pair content decreases the mass 
of the effective particles. 
Thus the greater the amount of pairs, the greater the acceleration, as shown by
the bottom panel in Fig.~\ref{fig:pairscold}.

\item {\it Hot cooling plasma -- No feedback ---} 
We consider a hot plasma with $\langle \gamma^2 \rangle = 10^6$ 
with cooling enabled. 
Fig. \ref{fig:pairshot} shows the force experienced by the layer (top panel), 
the variation of $\langle \gamma^2_{\rm L} \rangle$ (middle panel) and the evolution of 
$\left(\Gamma \beta\right)_{\rm L}$ (bottom panel) as a function of $z/R$.
As expected, hotter plasmas experience a stronger and negative initial 
force which tends to decelerate them. 
The decelerations experienced depend upon the pair content, thereby the pair dominated plasma 
is decelerated more than the others. 
This slowing down of the plasma leads to a decrease of the (negative)  force 
(due to the decrease of the received radiation energy density). 
This impacts the cooling rate as well as the force profile for the pair--dominated plasma. 
The lower cooling rate along with geometrical effect (entire spine irradiating the layer particle) 
is responsible for the burst in acceleration at large values of $z$.

% --------------------------------------
\begin{figure} 
\includegraphics[width=80mm]{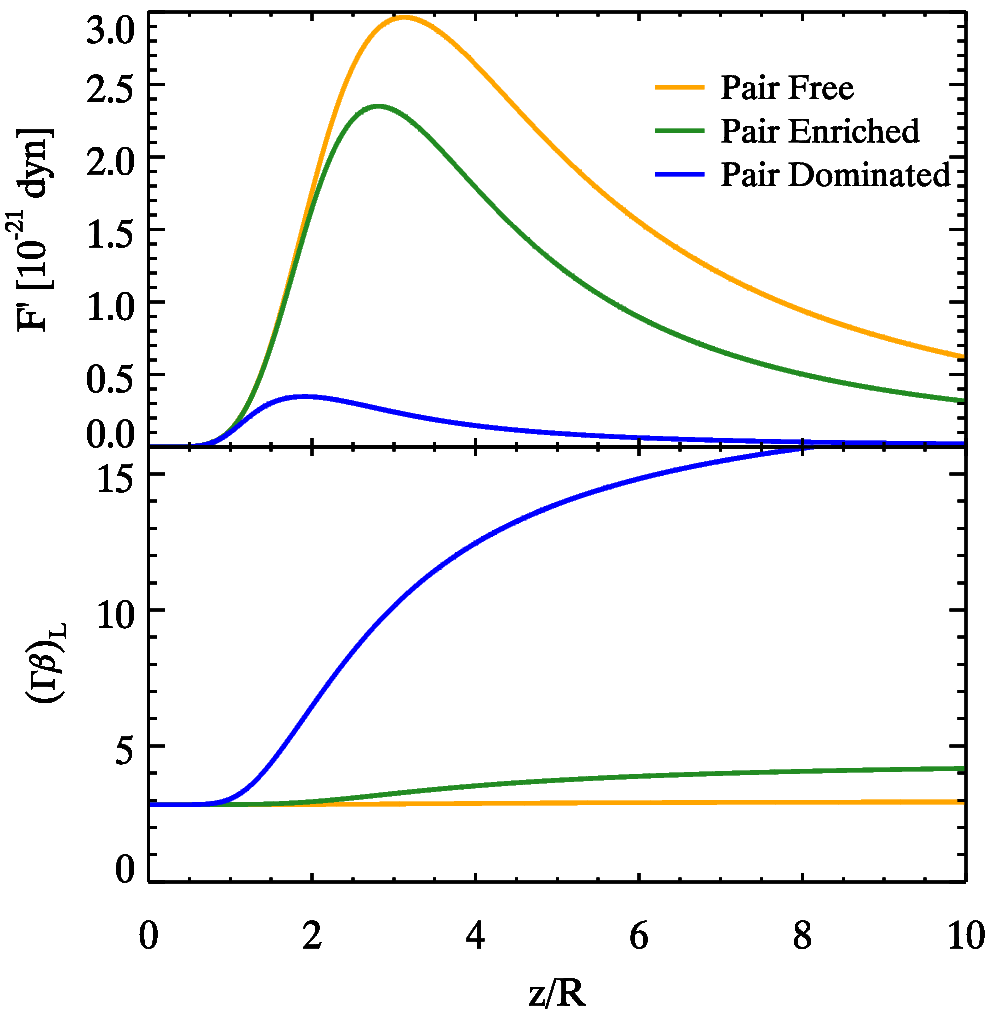}
\caption{
Radiative acceleration of a cold, pair--loaded layer plasma by varying lepton to proton ratio $f$; 
we analyze three cases with $\g2av=1$: no pairs in layer ($f=1$); 
a plasma with $f=20$ and the extreme case of a pair dominated plasma.
Parameters used: $\Gamma_{\rm S}=15$, $\Gamma_{\rm L,0}=3$ 
and $L^{\prime\prime}_{\rm S}=10\cdot L^{\prime}_{\rm L}$.
Top panel: radiative force as measured in the layer frame.
Bottom panel: profile of $\left(\Gamma \beta\right)_{\rm L}$ for the layer.
}
\label{fig:pairscold}
\end{figure} 
% --------------------------------------

% --------------------------------------
\begin{figure} 
\includegraphics[width=80mm]{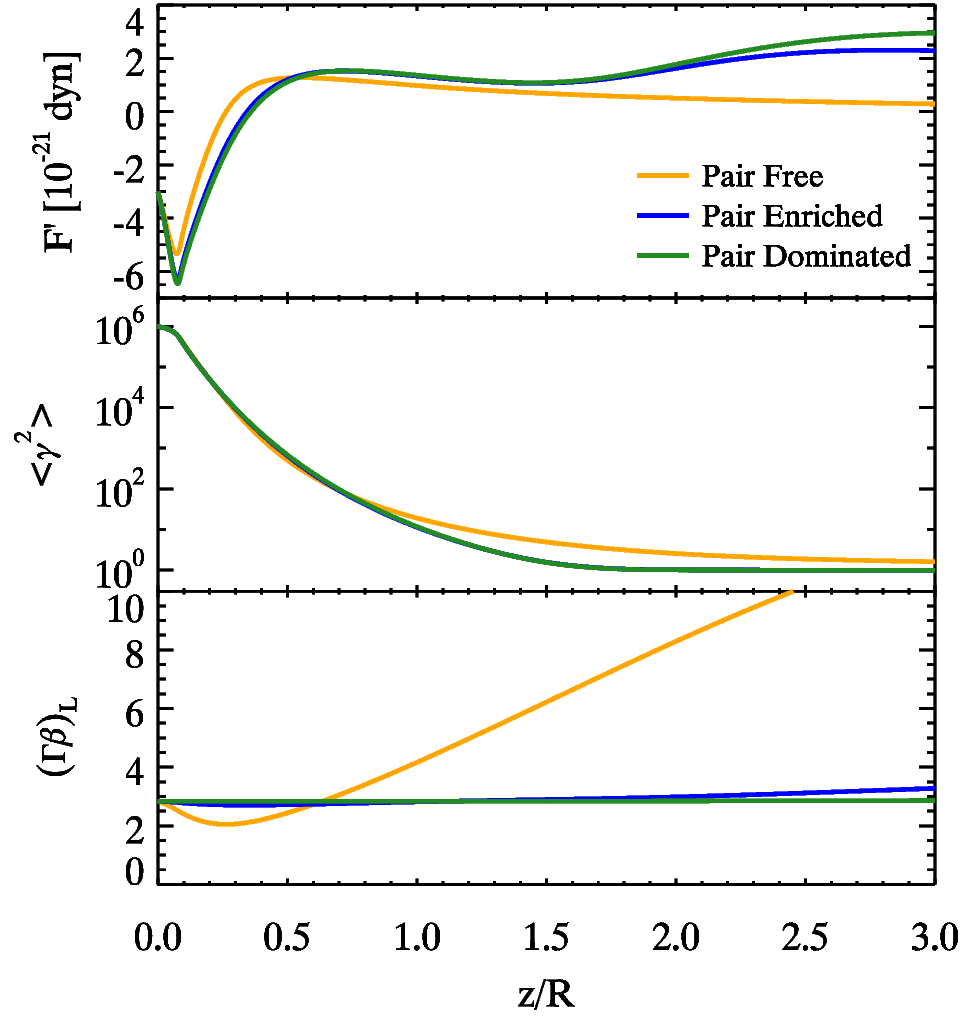}
\caption{
Radiative acceleration plot for a hot layer plasma having different 
lepton to proton ratios $f$; we analyze three distinct cases with an initial 
$\g2av_{\rm z=0}=10^6$: no pairs in layer and spine plasma ($f=1$);
a pair--enriched $f=20$ plasma, and the extreme case of a pair dominated plasma.
Parameters used: $\Gamma_{\rm S}=15$, $\Gamma_{\rm L,0}=3$ 
and $L^{\prime\prime}_{\rm S}=10\cdot L^{\prime}_{\rm L}$.
Top panel: radiative force measured in layer frame as a function of $z/R$. 
Middle panel: the evolution of internal energy of the sheath denoted by $\langle \gamma^2_{\rm L}\rangle$.
Bottom panel: profile of $\left(\Gamma \beta\right)_{\rm L}$ for the layer as a function of $z$.
}
\label{fig:pairshot}
\end{figure} 
% --------------------------------------

% --------------------------------------
\begin{figure} 
\includegraphics[width=80mm]{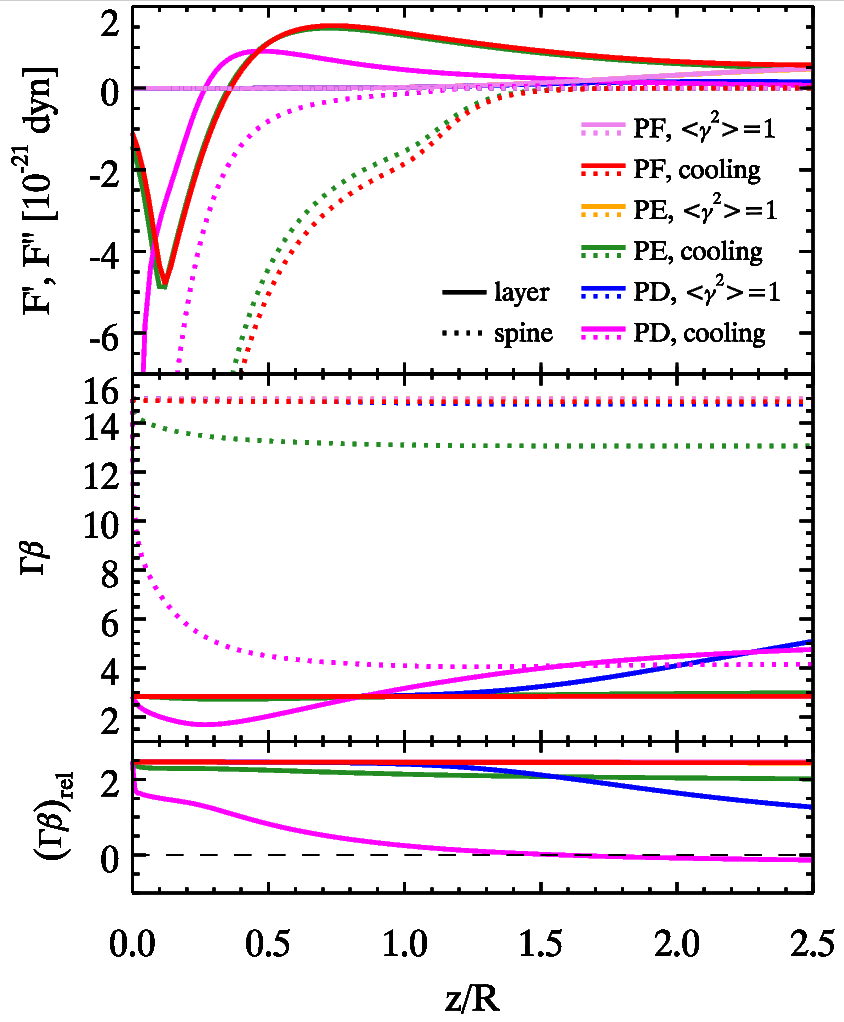}
\caption{
Radiative acceleration of the spine--layer pair loaded plasma with feedback; we compare and analyze three cases with different lepton to proton ratios: no pairs in the layer and spine plasmas (i.e. there is one lepton for each proton - PF); a plasma with $f=20$ leptons
for each proton (PE) and as the final case, a plasma dominated by pairs (PD).
Parameters used: $\Gamma_{\rm S, 0}=15$, $\Gamma_{\rm L,0}=3$ 
and $L^{\prime\prime}_{\rm S}=10\cdot L^{\prime}_{\rm L}$.
Top panel: Radiative force measured in spine (layer) frame in dotted 
(solid) lines as a function of $z/R$.
Middle panel: Profile of $\left(\Gamma \beta\right)$ for the spine (layer) 
represented by dotted (solid) lines. 
Bottom panel: The relative velocity profile of the spine with respect to the 
layer expressed as $\left(\Gamma \beta\right)_{\rm rel}$.
}
\label{fig:feed_pairs}
\end{figure} 
% --------------------------------------

\item {\it Feedback ---}
Now we discuss the third case, where we compare hot and cold plasmas with {\it both cooling and feedback enabled}. Fig.~\ref{fig:feed_pairs} depicts the evolution of the 
forces $F'$ and $F"$ (upper panel) as measured in the frame of the layer and of the spine, respectively, and $\left(\Gamma \beta\right)_{\rm L}$.
The forces $F'_{\rm S}$ experienced by the spine due to the radiative interaction with the 
layer are always negative leading to its deceleration. 
For cold plasmas (with initial $\langle \gamma^2 \rangle=1$) the deceleration, though small, 
is non--zero as compared to hot plasma with initial $\langle \gamma^2 \rangle=10^6$. 
We also note that with increasing $f$ values, the deceleration becomes much stronger.

Similar to the results of previous sections the layer force $F'_{\rm L}$ (denoted by the solid curves 
in the upper panel of Fig. \ref{fig:feed_pairs}) starts off negative and turns positive 
(due to geometrical effects). 
The initial magnitudes are proportional to both $\langle \gamma^2 \rangle$ and $f$. 
As a result the pair--dominated hot layer is initially significantly decelerated and then accelerates. 
At later times, %/ increasing values of $z$,
because of cooling the acceleration is no longer a function of $\langle \gamma^2 \rangle$ 
but still depends upon the particle ratio $f$. 
As a result, pair dominated $\Gamma_{\rm L, sat} \sim 6$ 
%- Francesco, could you read from the plot (I think we have the value in the older plot) the saturation value for the pair dominated plasma here} 
and enriched plasmas $\Gamma_{\rm L,sat} \sim 3.1 $ 
accelerate even for $z/R>5$ and much more than pair--free plasmas.
\end{enumerate}
% --------------------------------------

We can conclude that pair loading enables the plasma to accelerate significantly even 
when the plasmas are cold to begin with. 
Cold pair--enriched and the cold pair--dominated plasma exhibit significant acceleration, 
achieving layer Lorentz factors $\Gamma_{\rm L} \sim 4$ and $\Gamma_{\rm L} \sim 16$ respectively.

%Should we use the average $\Gamma$ or the exact profile?

%Using the average and $\langle \gamma^2_S\rangle =10^6$:
%$\Gamma_{\rm S}$ from 15 to 5....

%--- Cooling, feedback:  only to say that if cooling is on, there is no effect....

\section{Conclusions}
\label{sec:discnconclu}

In this work we have investigated the effects of radiative acceleration through Compton 
scattering in a simple case of structured jet: the spine--layer scenario. 
We summarize here the several factors we have explored in our work that influence 
the dynamical evolution of a structured jet.

There are two main acceleration regimes:
%  ---------------------------------
\begin{itemize}

\item[--] Compton Rocket Effect: For the Compton rocket to be effective the leptonic 
distribution must be hot ($\g2av \gg 1$). 
% i.e., there is significant random motion in the scattered plasma bulk reference frame. 
The seed photons do not directly provide the driving force but act as a catalyst for the 
accelerated motion, in fact the bulk kinetic energy is supplied by the internal energy of the plasma. 

\item[--] Radiatively Driven Motion/Normal Compton scattering: For this case the leptonic distribution 
of the plasma is cold ($\g2av\sim 1$).
%  i.e., there is no random motion in the scattered plasma's bulk  reference frame. 
The motion is driven solely by Compton scattering photons off cold electrons. 
The bulk kinetic energy of the plasma is supplied by the seed photon field due to momentum transfer to the plasma. 
The forces and hence the acceleration achieved in this case are small as compared to the Compton rocket effect.

\end{itemize}
%  ---------------------------------
Having identified the acceleration regimes, we will now summarize several factors important for acceleration:

%  ---------------------------------
\begin{itemize}

\item[--] Radiative Cooling : Radiative cooling decreases the internal energy content ($\g2av$) of the plasma 
quite rapidly which effectively kills the force -- thereby quenching the acceleration process. 
Thus it plays an important role in determining whether the Compton rocket effect or the radiatively 
driven motion dominates the kinematics of the structured jet. 
In \S \ref{subsec:nocool_nofeed} we discussed how an initially hot plasma experiences a decelerating force 
for small values of $z$ due to radiation directed along the negative $z$ direction or coming from spine 
regions located above the layer. 
%When radiative cooling is active the hot plasma cools rapidly while it simultaneously decelerates. 
As the plasma decelerates, it will simultaneously cool rapidly if radiative cooling is active.
The magnitude of this decelerating force decreases due to two reasons -- 
i) rapid cooling and ii) as the plasma travels further away from the base 
it receives some upward directed photons from the base (which is now to the 
rear of the layer plasma) pushing the layer plasma along the positive $z$ direction. 
As a result, the magnitude of the decelerating force rapidly decreases and it 
does not produce a significant change in $\Gamma_{\rm L}$ values.
As noted in \S~\ref{subsec:continj}, the radiative cooling activated by low or moderate 
spine luminosities quenches the acceleration process -- which effectively freezes the 
Lorentz factor of the layer at the value it had just before activation.

\item[--] Feedback: the feedback scenario enables us to study the self--consistent evolution 
of the spine and layer under the radiative influence of the layer and the spine respectively. 
Among the cases we studied, in the cooling regime neither the spine nor the layer shows any acceleration.
In some cases, the layer accelerates and even overtakes the spine -- 
but this requires a continuous injection of huge (and possibly unrealistic) amounts of energy. 
In the non--cooling or continuous energy injection regime, for $\g2av > 2 \times 10^4$, 
significant accelerations can be seen for the layer and decelerations for the spine. 
Interestingly, by making the situation more realistic by limiting energy injection 
(having a energy injection limited to the active region as explored in \S~\ref{subsec:continj} 
and \S~\ref{subsec:feedback:continj}) and by incorporating radiative cooling - the layer achieves 
Lorentz factors $\sim 3$ comparable to those required by observations. 
Thus radiative feedback induced acceleration plays in important role in regulating the bulk Lorentz 
factors of both the spine and the layer.

\item[--] Pair--loading:
If electron positron pairs are present, they decrease the effective mass of the plasma. 
As a result, the same forces (for non pair loaded plasmas) can accelerate pair loaded plasmas to higher Lorentz factors. Furthermore, even cold pair--loaded plasmas can attain high $\Gamma_{\rm L}$ values as 
compared to non pair--loaded plasmas, as shown in Fig.~\ref{fig:pairscold}.

\item[--] Factors influencing acceleration: the luminosity and the internal energy content 
of the plasma play a very important role in the acceleration process. 
In general, the greater the luminosity the greater the observed acceleration. 
The same is true for the amount of the internal energy that can be converted 
into bulk motion by the Compton rocket effect.
But this depends critically on the presence of same re--acceleration mechanism,
able to maintain the plasma hot.
% ame statement, i.e., the greater the internal energy content, the greater 
% the acceleration cannot be made, as shown in \S \ref{subsec:cool_nofeed} radiative cooling 
% can lead to an initially comparatively cooler plasma accelerating beyond an initially much hotter plasma. 
In several scenarios explored, the maximum of the accelerating force occurs outside the active or the standing 
shock region -- when the spine/layer particle observes the entire layer/spine irradiating it and 
thereby pushing it along the positive $z$ direction. 
This effect was first noticed while investigating the (rather unrealistic) 
no cooling scenario 
(see \S~\ref{subsec:nocool_nofeed} where cooling was permanently switched off). 
\end{itemize}
%  ---------------------------------
%\vskip 1 cm
%\vskip 1 cm
Regarding the possibility that the layer can be entirely accelerated by the 
radiative force, we can conclude that this is possible if the Compton rocket effect remains strong for a sufficiently long time, namely for the time needed to cross the active region $z\sim R$. This requires that $\g2av$ remain large ($\gsim 10^4$) within the layer, 
and this in turn demands continuous injection of fresh energetic leptons throughout the layer length.
Electrons--positron pairs can help, but are not crucial, since the maximum number
of pairs per proton is limited.

If the magnetic field within the layer is $B$, we expect that its synchrotron
emission peaks at $\nu_{\rm S, L} \sim  3.6\times 10^6 \gamma_{\rm b}^2 B\delta_{\rm L}$.
According to Eq. 6 (with $\gamma_{\rm max}\gg \gamma_{\rm b}$) we have 
$\gamma_{\rm b}^2 \sim  10^8  (\g2av/10^4)^2$ leading to
$\nu_{\rm S, L} \sim 3.6\times 10^{14} (\g2av/10^4)^2 B\delta_L$ Hz.
For $B\sim$1 G, similar to the magnetic field of the spine of blazars in the $\gamma$--ray
emitting zone (Tavecchio \& Ghisellini 2015),
we have that layers accelerated radiatively should peak in the optical--UV
band, and there should be a relation between their synchrotron peak frequency and
their bulk Lorentz factor.
The higher $\nu_{\rm S, L}$ the larger $\Gamma_{\rm L}$, and the smaller the 
relative $\Gamma$ between the spine and the layer.

On the contrary, if $\nu_{\rm S, L}$ is small, then $\g2av$ is also small, suggesting
$\Gamma_{\rm L }\sim 1$. 
Although the relative $\Gamma$ approaches $\Gamma_{\rm S}$, the radiative interplay 
between the two structures should be weak, since the layer cannot produce 
many seed photons if its $\g2av$ is small.

There is therefore a defined range of $\nu_{\rm S, L}$ where radiative acceleration of the
layer can work.
If $\nu_{\rm S, L}$ is in the far infrared and we have indications that $\Gamma_{\rm L}\sim 3$ or more,
then it is very likely that the layer was not accelerated radiatively, but by the same process that accelerated the spine.
The model can thus be tested studying the spectral energy distribution
of blazars and radio--galaxies.
The blazars where we can reliably derive the spectral parameters of the layer
are still too few to draw any strong conclusions.
For radio--galaxies, we should be careful to select those whose observed emission is reliably associated to a layer located in the inner region of the relativistic jets, and not to more extended components. So, the selected sources should show rapid variability indicating a compact emitting region.
%So we should select sources whose rapid variability indicates a compact emitting region.

We have thus shown how structured spine--sheath jets are radiatively accelerated in the Compton 
rocket and radiatively driven motion regimes. 
We have considered different values of Lorentz factors, 
luminosities and internal energy contents to understand the details of the acceleration process and have been successful in developing some insight and intuition regarding the phenomena. We have also shown that by including radiative feedback (between the spine and the layer), radiative cooling and a realistic energy injection model (the continuous injection scheme within the active region) the observed Lorentz factor of the layer can be reproduced. 
We have also proposed tests for our model by utilizing layer associated emissions from inner parts of radio-galaxy jets. The range of our obtained Lorentz factors predict that radiatively accelerated layers' synchrotron peak values  $\nu_{\rm S, L}$ occur in the IR-optical band.

% Interestingly, several considered cases achieve the desired saturation 
% of the layer bulk Lorentz factor $\Gamma_{\rm L} \sim 3$ or a few, such as the \S \ref{subsec:continj} 
% and \S \ref{subsec:pairsfeed} (pair dominated plasma in Fig. \ref{fig:feed_pairs}). 
% {\bf In most cases you start with $\Gamma_{\rm L}=3$ and, so ending with $\Gamma_{\rm L} \sim few$ 
% is not a very strong result! -- DL}

\section*{Acknowledgments}
GG and FT acknowledge contribution from a grant PRIN--INAF--2014. AC thanks and appreciates GG and INAF, Brera for their hospitality at Merate, IT.

\appendix
\section{Details of force calculation}

%In this section we provide a detailed calculation of the force experienced by a layer 
%element at a position $z$. The equations below are a continuation of the expressions obtained in ~\ref{subsec:eqmotion}. 
Consider an element of the layer at position $z$; we call ${\rm d}U^\prime$ %(\theta)$
%, \Gamma_{\rm L}, \Gamma_{\rm s})$
the differential energy density of the spine radiation received from 
an angle between $\theta$ and $\theta + {\rm d}\theta$, 
measured in the frame of the layer. 
The contribution of this radiation energy density to the infinitesimal force ${\rm d}F^\prime_{\rm z}$ parallel to the jet axis and acting on the 
layer particle is given by:
\begin{equation}
 {\rm d}F^\prime_{\rm z} \, = \, \frac{16}{9}\sigma_{\rm T} \langle \gamma^2 \rangle \cos \theta^\prime {\rm d}U^\prime %(\theta)
 \label{eq:forza1}
\end{equation}
where $\theta^\prime$ is the angle of incoming photons 
with respect to the jet axis direction as seen by the layer.
The total force exerted on the layer's effective particle can be computed by
integrating the Eq. \ref{eq:forza1} over the incoming angles $\theta^\prime$.
Since the spine is active between points that are fixed in the observer 
frame $K$, it is easier to compute the integral in that frame $K$.
From the relations of aberration of the light 
(e.g. Weinberg, 1972) we have useful transformations:
\begin{equation}
\cos \theta^\prime \, = \,  \frac{\cos \theta-\beta_{\rm L}}{1-\beta_{\rm L}\cos \theta}
\label{eq:abercos} 
\end{equation}
\begin{equation}
{\rm d}\Omega^\prime \, = \,  {\rm d}\Omega \cdot \delta_{\rm L}^2
\label{eq:domega}
\end{equation}
The differential radiation energy density  
can be written as:
\begin{equation}
{{\rm d}U^\prime \over {\rm d}\Omega^\prime} \, = \,  \frac{I^\prime}{c}
\label{eq:du1}
\end{equation}
where $I^\prime$ is the bolometric radiation intensity 
as seen by the layer and it is related to the spine 
comoving radiation intensity $I^{\prime\prime}$ by:
\begin{equation}
I^\prime \, = \, I^{\prime\prime} \cdot \delta^4_{\rm S,L}
\label{eq:ip}
\end{equation}
Consider now that the uni--dimensional spine is actually 
an infinitesimal cylinder whose axis is coincident with 
the jet axis of height $R$ and radius $r \rightarrow 0$. 
In this case we have: 
\begin{equation}
I^{\prime\prime}\, = \, j^{\prime\prime} \cdot r \, = \, \lambda^{\prime\prime}_{\rm S} \frac{1}{\pi r}
\label{eq:ipp}
\end{equation}
where $j^{\prime\prime}$ is the comoving spine emissivity. 
$\lambda^{\prime\prime}_{\rm S} = \left[\dfrac{{\rm d}L_{\rm S}^{\prime\prime}}{{\rm d}x^{\prime\prime}}\right]$ 
is the comoving spine luminosity linear density profile; 
it is generally a function of the position $z$, 
but in case of uniform luminosity distribution we can write:
\begin{equation}
\lambda^{\prime\prime}_{\rm S}=\dfrac{{\rm d}L_{\rm S}^{\prime\prime}}{{\rm d}x^{\prime\prime}} \, = \, {L_{\rm S}^{\prime\prime} \over R^{\prime\prime}} \, = \, {L_{\rm S}^{\prime\prime} \Gamma_{\rm S}\over R} 
\label{eq:proflum}
\end{equation}
where $R^{\prime\prime}=R/\Gamma_{\rm S}$ is the length 
of the spine active region measured in the frame 
comoving to the spine. 
$L_{\rm S}^{\prime\prime}$ is the total comoving luminosity 
of the spine and it is related to the observed isotropic 
luminosity by (see Lind \& Blandford, 1985): 
\begin{equation}
L_{\rm S,iso} \, = \, L_{\rm S}^{\prime\prime} \delta^3_{\rm S}(\theta_{\rm view}) \, = \, \dfrac{ L_{\rm S}^{\prime\prime}}{\Gamma^3_{\rm S}(1-\beta_{\rm S}\cos\theta_{\rm view})^3}
\label{eq:liso}
\end{equation}
where $\theta_{\rm view}$ is the angle between the jet axis and the line of sight.
That relation is different from the usual one (that requires a factor $\delta^4$
between the rest frame and the isotropic luminosity) because in this case there the rest frame emitting volume properly is ill-defined as the end points of the emitting region are moving with respect to the emitting fluid.
We can relate the observed luminosity density $\lambda_{\rm S}$
with the comoving one through Eq. \ref{eq:comdens}:
\begin{equation}
\lambda_{\rm S} \, = \, \dfrac{L_{\rm S,iso}}{R} \, = \,  \lambda^{\prime\prime}_{\rm S}\dfrac{\delta^3_{\rm S}(\theta_{\rm view})}{\Gamma_{\rm S}}
\label{eq:comdens}
\end{equation}
Using the definition of solid angle ${\rm d}\Omega$ we can write:
\begin{equation}
{\rm d}\Omega \, = \, \frac{{\rm d}A}{D^2} \, = \, \frac{{\rm d}x \cdot 2r}{D^2} \, = \, \frac{{\rm d}\theta \cdot 2br}{D^2\sin^2 \theta} \, = \, \frac{2r}{b}{\rm d}\theta
\label{eq:dom}
\end{equation}
where ${\rm d}A=2r{\rm d}x$ is the differential area 
of the spine seen under the angle ${\rm d} \Omega$, 
$D=b/\sin \theta$ is the distance between the emitting element 
of the spine and the layer, $x=-D\cos\theta=-b \cot\theta$ 
is the projection of $D$ over the jet axis.
Using eqs. \ref{eq:deltasl}, \ref{eq:domega}, \ref{eq:du1}, \ref{eq:ip}, 
\ref{eq:ipp}, \ref{eq:dom}, we can write:
\begin{equation}
{{\rm d}U^\prime (\theta)\over {\rm d}\theta} \, = \,  \lambda^{\prime\prime}_{\rm S} \eta \dfrac{\delta^4_{\rm S}}{\delta^2_{\rm L}}\dfrac{1}{bc}
\label{eq:du2}
\end{equation}
that depends no more on $r \rightarrow 0$. In Eq. \ref{eq:du2}, $\eta$ is a factor of the order of unity that depends on the geometry (in our case we use $\eta=2/\pi$).
The total force exerted on the layer test particle 
is obtained integrating Eq. \ref{eq:forza1}:
\begin{equation}
F^\prime_{\rm z} 
\, =  \, \frac{16}{9}{\sigma_{\rm T} \over bc} \langle \gamma^2 \rangle \eta \int_{\theta_1}^{\theta_2} \lambda^{\prime\prime}_{\rm S}  \dfrac{\delta^4_{\rm S}}{\delta^2_{\rm L}}\dfrac{\cos\theta-\beta_{\rm L}}{1-\beta_{\rm L}\cos\theta}{\rm d}\theta
\label{eq:totalforce}
\end{equation}
The limits of integration $\theta_1$ and $\theta_2$ 
are measured in the central engine frame 
(comoving to the limits of the emitting volume) 
and depend only on the position of the layer $z$:
\begin{equation}
\cos\theta_1 \, = \, \frac{z}{(b^2+z^2)^{1/2}}
\label{eq:costh1}
\end{equation}
\begin{equation}
\cos\theta_2 \, = \, \frac{z-R}{(b^2+z^2+R^2-2Rz)^{1/2}}
\label{eq:costh2}
\end{equation}
Evaluating the force at each value of $z$ we can obtain the layer Lorentz factor profile $\Gamma_{\rm L}(z)$ by numerically solving Eq. \ref{eq:dgb}.

\end{document}